\documentclass[footinbib, aps,prl,reprint, 
10pt,twocolumn,showpacs,showkeys]{revtex4-1}
\usepackage{amsfonts,amsmath,amssymb}
\usepackage{blindtext}
\usepackage{graphicx}
\usepackage[pdfstartview=FitH,colorlinks=true,linkcolor=blue,citecolor=blue,urlcolor=blue]{hyperref}
\usepackage{tikz}
\usepackage{verbatim}

\usepackage{times}

\usepackage[english]{babel}
\usepackage{latexsym}
\usepackage{graphics}
\usepackage{subfigure}
\usepackage{epsfig}
\usepackage{color}
\usepackage{braket}
\usepackage[T1]{fontenc}
\usepackage[latin9]{inputenc}
\usepackage{amstext}
\usepackage{esint}
\usepackage[normalem]{ulem}   
\usepackage{subfigure}

\newcommand{\beq}{\begin{equation}}
\newcommand{\eeq}{\end{equation}}

\begin{document}

\title{Equation of state of Bose gases beyond the universal regime}

\author{Mart\'\i \ Planasdemunt}
\affiliation{Departament de F\'isica, Universitat Polit\`ecnica de Catalunya, Campus Nord B4-B5, 08034 Barcelona, Spain}

\author{Jordi Pera}
\email{jordi.pera@upc.edu}
\affiliation{Departament de F\'isica, Universitat Polit\`ecnica de Catalunya, Campus Nord B4-B5, 08034 Barcelona, Spain}

\author{Jordi Boronat}
\email{jordi.boronat@upc.edu}
\affiliation{Departament de F\'isica, Universitat Polit\`ecnica de Catalunya, Campus Nord B4-B5, 08034 Barcelona, Spain}

\date{\today}

\begin{abstract}
The equation of state of dilute Bose gases, in which the energy 
only depends on the $s$-wave scattering length, is rather unknown beyond the 
universal limit. We have carried out a bunch of diffusion Monte Carlo 
calculations up to gas parameters of $10^{-2}$ to explore how the departure 
from the universality emerges. Using different model potentials, we calculate 
the energies of the gas in an exact way, within some statistical noise, and 
report the results as a function of the three relevant scattering parameters: 
the $s$-wave scattering length $a_0$, the $s$-wave effective range $r_0$, and 
the 
$p$-wave scattering length $a_1$. If the effective range is not large we 
observe universality in terms of $a_0$ and $r_0$ up to gas parameters of 
$10^{-2}$. If $r_0$ grows the regime of universality in these two parameters is 
reduced and effects of $a_1$ start to be observed. In the $(a_0,r_0)$ 
universal regime we propose an analytical law that reproduces fairly well the 
exact energies. 
\end{abstract}

\maketitle

The experimental realization of the Bose-Einstein condensed state (BEC) 
\cite{bec1,bec2} has 
allowed for the study of a dilute Bose gas whose main properties were predicted 
theoretically many years ago. The high tunability of the interaction strength 
in these systems offers a unique platform to gain accurate insight on their 
properties. However, this intriguing feature cannot be carried out  as far as 
theory would like due to the metastability of the dilute BEC in the 
laboratories \cite{giorgini_rmp} . Three-body losses frequently hinder the 
achievement of large 
densities $n$ and universality in terms of the gas parameter $x = n a_0^3$, 
with 
$a_0$ the $s$-wave scattering length, is what is normally observed. 
Nevertheless, we already know of experimental results that go beyond
universality: the Fermi polaron in two dimensions \cite{Darkwah2019,bombin1}, 
dilute Bose-Bose liquid drops \cite{tarruell,viktor}, and dipolar gases 
\cite{pfau,bombin2}.

The first terms of the equation of state of Bose gases were calculated long 
time ago and proved to be dependent only on the gas parameter,
\begin{equation}
\frac{E_{\text u}}{N} = c_{\text{MF}} \, x + c_{\text{LHY}}\, x^{3/2} + 
c_{\text{WU}} \, x^2 
\ln x \ ,
\label{MFLHY}
\end{equation}
with the energy in unities of $\hbar^2/(2 m a_0^2)$. The universal constants in 
Eq. (\ref{MFLHY}) are
\begin{eqnarray}
c_{\text{MF}} & = & 4 \pi \ ,\\
c_{\text{LHY}} & = & \frac{512 \sqrt{\pi}}{15} \ , \\
c_{\text{WU}} & = & \frac{32 \pi (4 \pi-3 \sqrt{3})}{3} \ .
\label{coefMF}
\end{eqnarray}
The first term in Eq. (\ref{MFLHY}) is the mean-field or Hartree-Fock 
term~\cite{bogoliubov}. The second one was first calculated by Lee, Huang, and 
Yang 
\cite{leehuangyang} and incorporates quantum fluctuations at the lowest order, 
and 
finally the last one was obtained by Wu~\cite{wu}. The LHY and WU terms were 
first derived by a hard-sphere interaction and, later on, it was proved that 
they are also valid for repulsive potentials with the same $s$-wave 
scattering 
length~\cite{sawada1,sawada2,beliaev,lieb}. Beyond $E_{\text{u}}/N$ 
(\ref{MFLHY}), the energy of the 
Bose gas is no more a universal expression depending only on the $s$-wave 
scattering length of the potential. Hugenholtz and Pines~\cite{hugenholtz} 
proposed a 
continuation of the series (\ref{MFLHY}) in the form
\begin{equation}
\frac{E}{N} = \frac{E_{\text u}}{N} + d_1 x^2 + d_2  x^{5/2} \ln x + \ldots \ ,
\label{Pines}
\end{equation}
with new coefficients $d_i$ which must depend on the shape of the interatomic 
potential. These parameters are expected to be related to scattering parameters 
corresponding to a larger momentum transfer, essentially the $s$-wave effective 
range $r_0$ and the $p$-wave scattering length $a_1$, and also to three-body 
$s$-wave contributions \cite{braaten,braaten2,braaten3,hammer}. This 
perturbative series for a dilute Fermi gas is well 
settled up to third order of the gas parameter, the terms beyond the universal 
regime being known, some of them with analytic expressions and the rest with 
integrals that are numerically estimated \cite{bishop,pera2024}. In this 
expansion for fermions, the 
analytic terms depend explicitly on $r_0$ and $a_1$. The knowledge of the 
nonuniversal terms for the equation of state of a Bose gas is very reduced up 
to now due to difficulties in the calculation of the required diagrams. The 
most recent attempt to continue the perturbative series was carried out by 
Braaten \cite{braaten,braaten2}. In this work, an expression beyond the LHY 
term is proposed with terms 
that correspond to three-body contributions and another one that depends on 
$r_0$. The 
weight of the three-body term was determined by a fitting procedure to 
diffusion Monte Carlo (DMC) results available at that time \cite{boronat}. As 
these data was 
focused to low densities, this parameter was determined with a large 
uncertainty.

\begin{figure*}[t]
    \centering
    \includegraphics[width=0.9\textwidth]{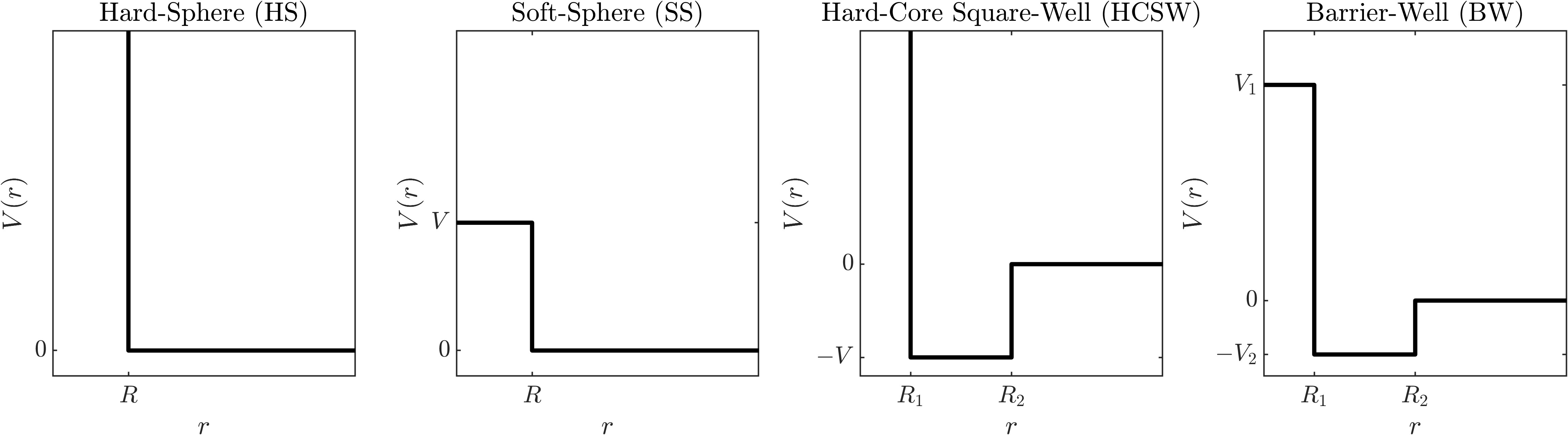}
    \caption{Model potentials used in the DMC calculations.}
    \label{potencials}
\end{figure*}

In this paper, we study the equation of state of Bose gases beyond the 
universal limit using the DMC method \cite{borodmc}. Using different model 
potentials (Fig.~\ref{potencials}), from 
which we know the low-energy scattering parameters, we explore an extended 
universality, but now in terms of $a_0$, $r_0$, and $a_1$. Our results show 
that this hypothesis is approximately correct within a certain domain of $r_0$ 
values and maximum density and that results with different interactions, 
sharing 
common scattering values, are always closer. We discuss the limitations of the 
Braaten series \cite{braaten} to reproduce the DMC results, mainly when the 
density grows. We 
propose an empirical form that fits well all the DMC data for a limited set of 
$|r_0| \lesssim 2$ values and gas parameters up to $1\times 10^{-2}$.

The DMC method solves stochastically the imaginary-time Schr\"odinger equation 
providing exact results for the ground-state energy and other properties of the 
system, within some statistical noise \cite{borodmc}. To reduce the variance 
one introduces a 
trial wave function that acts as importance sampling, leading the sampling to 
regions where the exact wave function is expected to be large and importantly 
eliminating divergences when the potential becomes singular. The model that we 
use is a standard Jastrow wave function $\Psi_{\rm J} ({\bf R})=\prod_{i<j}^N 
f(r_{ij})$, with $f(r)$ a two-body correlation function. We build $f(r)$ as the 
two-body solution, properly symmetrized to account well with the periodic 
boundary conditions (see specific details 
in Ref.~\cite{suplement}). The time-step 
and the average number of configurations (\textit{walkers}) is adjusted to 
eliminate any systematic bias in the results. It is worth noting that our 
Green's function is accurate to second order in the time step, reducing in this 
way the time-step dependence and allowing for the use of larger values which 
reduce the correlation length. The model potentials used in our simulations are 
shown in Fig.~\ref{potencials} and their characteristic values are chosen 
to produce the desired scattering parameters.

Our simulations work with finite numbers of particles, confined in a cubic box 
of the proper size to get the expected bulk density, and thus it is important 
to study any finite-size effect. This is particularly important in our case 
because we need to work with the largest possible accuracy. For a fixed density 
and interaction, we carry on a set of simulations with increasing number of 
particles, typically in the range $N=50$--$700$. Then, we get the energy in the 
thermodynamic limit $E_0/N$ using the linear law $E/N = E_0/N + \alpha \, 1/N$ 
\cite{suplement}. The time step and asymptotic population of 
walkers are chosen 
to eliminate any significant bias coming form these two technical parameters.   
   
Scattering theory at low energy, and up to momentum transfer $~k^2$, can be 
developed using three scattering parameters: the $a_0$ and $a_1$ scattering 
lengths~\cite{pera},
\begin{equation}
 a_l^{2l+1} = \frac{1}{2l+1}\int_0^\infty V(r)r^{l+1}u_l^{(0)}(r) dr \ ,
 \label{scata}
\end{equation}
corresponding to $l=0$ and $l=1$, respectively, and the $s$-wave effective range
\begin{equation}
r_0  = \frac{2}{a_0^2}\int_0^\infty   \left [ (r-a_0)^2 - 
{u_0^{(0)}}^2(r) \right ] dr   \ .
\label{erange}
\end {equation}
In Eqs. (\ref{scata},\ref{erange}), $V(r)$ is the interaction potential, 
$l$ is the angular momentum quantum 
number, and $u_l^{(0)}$ is the reduced radial wave function, solution of the 
radial two-body Schr\"odinger equation for that potential, with zero energy 
and angular momentum $l$. For the model potentials used in our work, these 
three 
parameters are analytically known as a function of the parameters defining the 
interaction~\cite{pera}. The inverse problem, that is, to know the potential 
parameters for 
a given set of scattering parameters is not trivial due to the complex 
character of the equations to solve. To this end, we implemented a stochastic 
search that proved to be very efficient (see Ref. \cite{suplement}).

\begin{figure*}[t]
    \centering
    \includegraphics[width=0.47\textwidth]{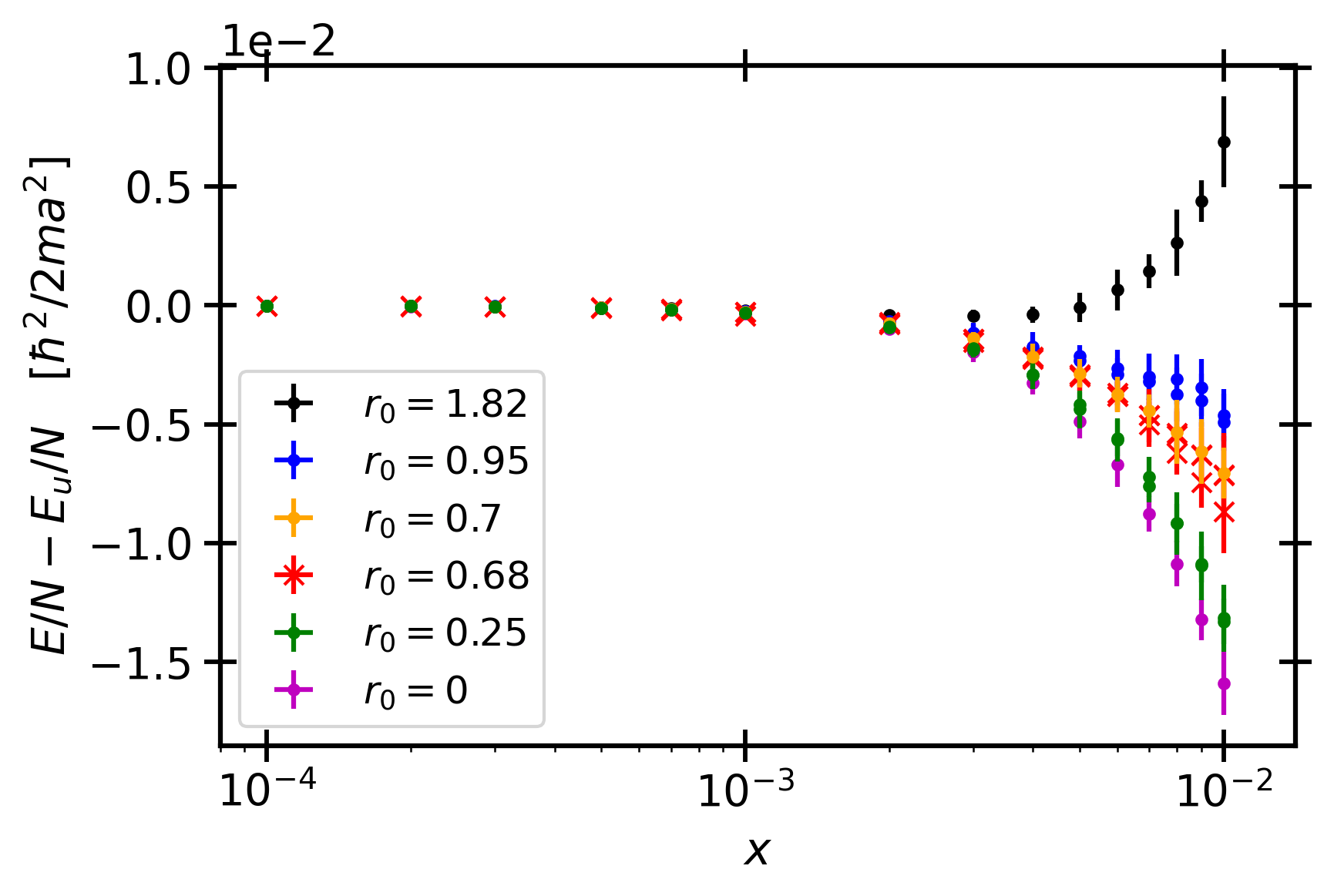}
      \includegraphics[width=0.47\textwidth]{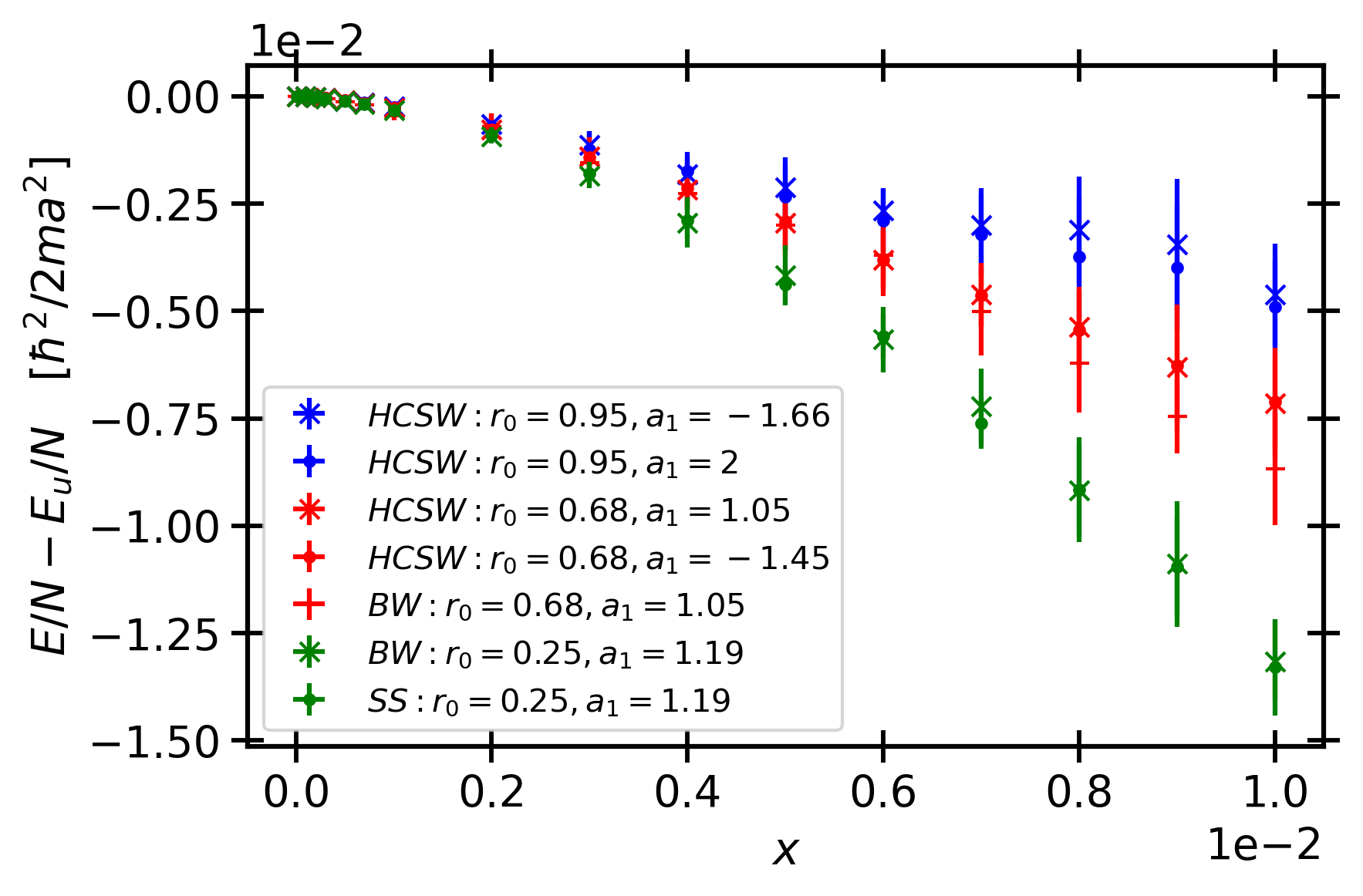}
 \caption{\textit{Left}: DMC energies using different model potentials with a 
common value of the $s$-wave effective range $r_0$ and $s$-wave scattering 
length $a_0$. The 
scattering length is used as unite length everywhere. To show the departure 
from the universal regime, we subtract to the energy per particle the 
universal term (\ref{MFLHY}). \textit{Right:} DMC energies using different 
potentials for values of $|r_0| < 2$. Same colors stand for common $r_0$ 
values.}
    \label{energies_dmc}
\end{figure*}

In Fig. \ref{energies_dmc} (left panel), we show the departure from the 
universal regime plotting the DMC energies using the universal energy 
(\ref{MFLHY}), with $c_{\text{WU}}=0 $, as a reference. From here on, we do not 
include the logarithmic Wu contribution because it presents a pathological 
behavior when $x$ grows, even below the universal limit~\cite{boronat}.
We have studied 
densities from $na^3=10^{-4}$ up 
to $10^{-2}$ and the potentials shown in Fig. \ref{potencials}. Below 
$~10^{-3}$, 
the differences between the different calculations are tiny and not visible at 
the scale of the plot. This result agrees with previous estimations obtained 
using also DMC \cite{boronat}, but now the accuracy is improved by using modern 
CPU's that allow 
for a more accurate determination of finite-size effects. Increasing more the 
density, one can appreciate as the energies spread and, at the larger value 
analyzed in our work ($10^{-2}$), the effect of finite-range effects is clear. 
A first significant finding of our study is that energies appear ordered from 
low to high values of $r_0$. To better analyze a possible universality in the 
($a$,$r_0$) plane, we show in Fig.  \ref{energies_dmc} (right panel) a zoom 
in a regime of $r_0$ values around one. Organized in different colors, we show 
the energies obtained using different potentials that share a common $r_0$ 
value. As one can see, at least in this window of $r_0$ values, the 
universality in terms of the two $s$-wave scattering parameters is clearly 
observed up to the largest density here explored. However, as we will discuss 
later on, the range of two-parameters universality is significantly reduced when 
$r_0$ becomes larger. The role of the $p$-wave scattering length in the cases 
reported in Fig. \ref{energies_dmc} (right panel) is quantitatively below our 
error bars, since in the figure we plot energies with the same $r_0$ but 
distinct 
$a_1$.

Braaten \cite{braaten} proposed a continuation of the perturbative series 
beyond the universal 
regime (\ref{MFLHY}) in the form
\begin{eqnarray}
    \frac{E}{N} -  \frac{E_{\text u}}{N} & = & 4 \pi x
\left[ \left( 
c_0+\frac{4\pi-3\sqrt{3}}{6\pi}\ln \bar{x}  \right) \bar{x}  \right. 
\label{eqbraaten} \\
& + &  \left.   \left(\frac{ 
8\left( 4\pi-3\sqrt{3}\right) 
}{3\pi^2} \ln \bar{x} +c_1-\frac{16}{15\pi}r_0\right) {\bar{x}}^{3/2}  \right] 
\nonumber \ ,
\end{eqnarray}
with $\bar{x}=16\pi x$. The parameter $c_0$, termed $c_E$ in 
Ref.~\cite{braaten}, is the three-body contact parameter. Braaten \textit{et 
al.} obtained the value of $c_0$ by fitting Eq. (\ref{eqbraaten}) to DMC 
energies of Ref. \cite{boronat}, but the scarce data available and their 
relatively large statistical errors made difficult its estimation. The results 
found were dependent on the specific interaction, for instance for a HS 
potential they obtained $c_0=-0.9 \pm 0.5$ and for a SS one $c_0=-1.8\pm 0.8$. 
It is worth noticing that our results for $c_0$ are compatible with the 
estimations in Ref.~\cite{braaten}, within the statistical errors.

\begin{figure*}[t]
    \centering
    \includegraphics[width=0.47\textwidth]{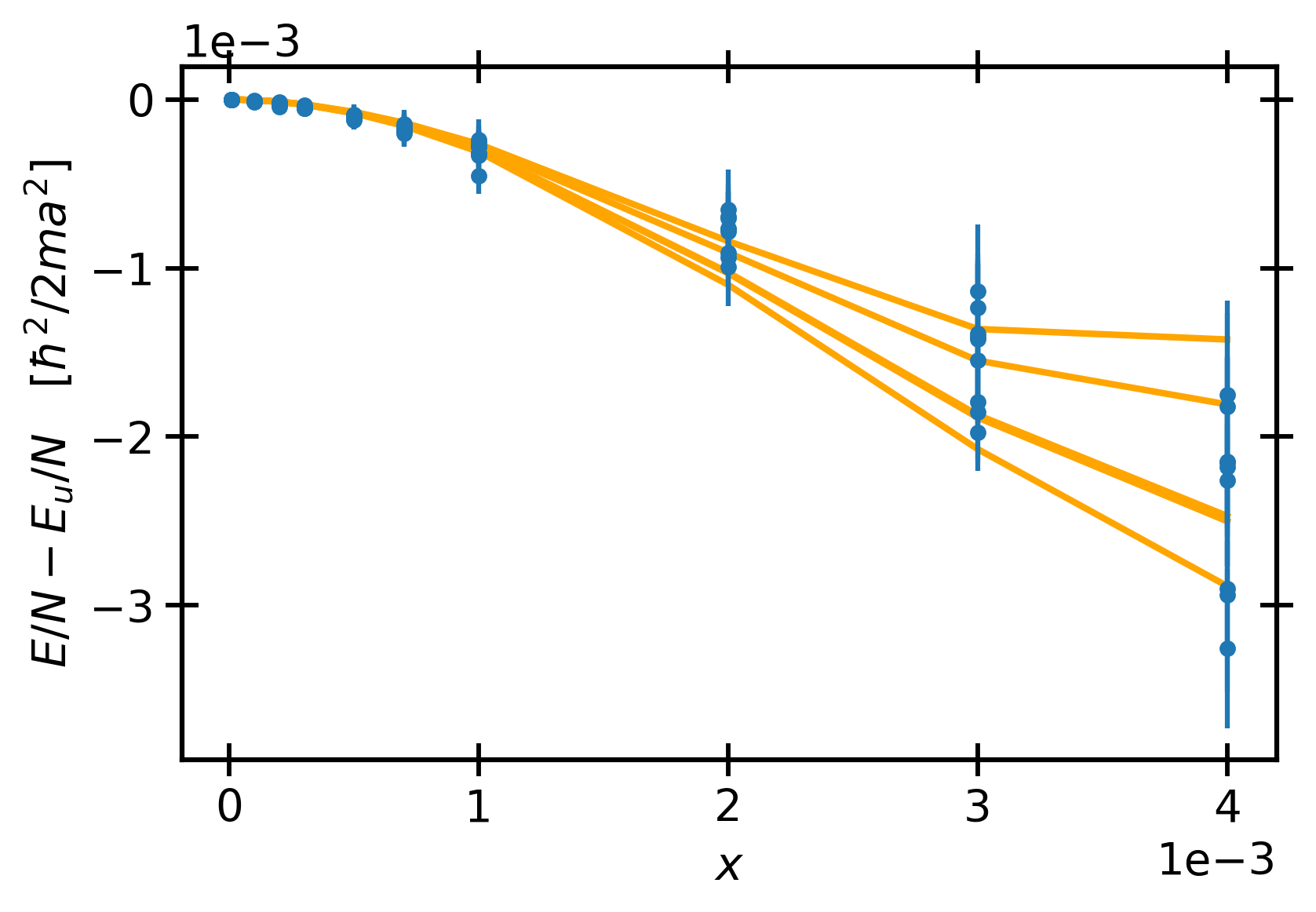}
      \includegraphics[width=0.47\textwidth]{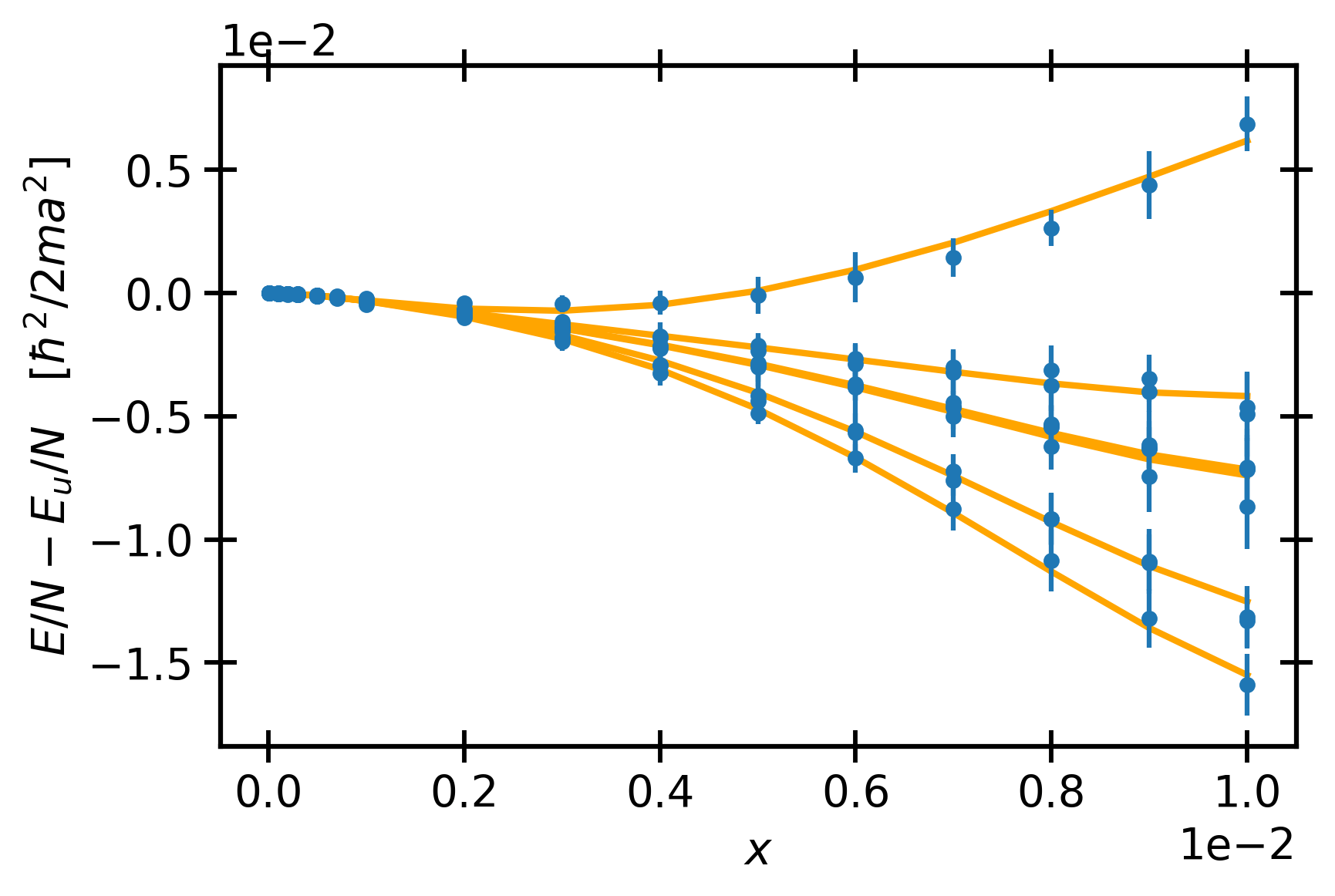}
 \caption{\textit{Left}: Braaten equation of state (\ref{eqbraaten}) and the 
set of DMC energies used for the fit. 
\textit{Right:} Our proposal for the equation of state (\ref{eqnostra}) and 
the DMC energies included in the fit. The best parameters are reported in Table 
\ref{tablenosaltres}.}
    \label{eos_dmc}
\end{figure*}

With the new data obtained in the present work we have repeated the Braaten 
analysis. We selected energies for effective ranges $|r_0| < 2$ and used both 
the original model (\ref{eqbraaten}) and the one in Eq. (\ref{eqbraaten}) but 
without the logarithmic terms. In both cases,  we get statistically compatible 
fits ($\chi^2/\nu \lesssim 1$) only for gas parameters $x \leq 4 \times 
10^{-3}$. With the complete Braaten model (\ref{eqbraaten}), we obtained 
$c_0=2.20 \pm   0.06 $ and $c_1=0.57 \pm 0.16 $ whereas, without the 
logarithmic terms $c_0= -0.73 \pm 0.06  $ and $c_1= 1.33 \pm 0.15 $, the latter 
showing 
a slightly better $\chi^2/\nu$. The three-body contact parameter $c_0$ in Eq. 
(\ref{eqbraaten}) keeps a rather constant value $c_0= 1.2 \pm 0.1$ for energies 
corresponding to gas parameters $x \leq 7\times 10^{-4}$. Its absolute value is 
close to the original Braaten estimation but the sign that we obtain is changed.
In Fig.~\ref{eos_dmc}, we show the better Braaten equation (\ref{eqbraaten}), 
restricted to the density regime where it is compatible with our results, and 
the set of DMC points used for the fit (see the table of all the DMC energies 
in Ref.~\cite{suplement}).

\begin{table}[b]
\centering
\caption{Set of optimal parameters of Eq. (\ref{eqnostra}). }
\begin{tabular}{ |c|c| } 
 \hline
 $c_0$ &  $-1.29\pm0.07$ \\ 
 \hline
 $c_1$ &  $5.4\pm0.4$ \\ 
 \hline
 $c_2$ &  $-9.9\pm0.9$ \\ 
 \hline
 $c_3$  & $6.2\pm0.6$ \\ 
 \hline
 $c_4$  & $2.5\pm0.08$ \\ 
 \hline
 $c_5$ &  $-2.33\pm0.12$ \\ 
 \hline
 \end{tabular}
 \label{tablenosaltres}
\end{table}

We extended our DMC calculations up to gas parameters $x=10^{-2}$, a value 
which clearly goes beyond the validity of Eq. (\ref{eqbraaten}). Inspired by 
the series expansion proposed by Hugenholtz and Pines~\cite{hugenholtz} 
(\ref{Pines}) 
we found an empirical equation of state in the form
\begin{eqnarray}
    \frac{E}{N} -  \frac{E_{\text u}}{N} & =  & 4 \pi x \left[ 
c_0 \bar{x} + \left(c_1-\frac{16}{15\pi}r_0 \right)  
{\bar{x}}^{3/2}   \right.  
  \label{eqnostra} \\
 & & \left.  + \left(c_2+c_4 
r_0\right)  {\bar{x}}^{2}  + \left(c_3+c_5 r_0\right)   
{\bar{x}}^{5/2} \right] \ ,\nonumber
\end{eqnarray}
which reproduces well our results. Compared with Eq (\ref{eqbraaten}), we 
notice that we have suppressed the logarithmic terms and increased the order of 
the expansion up to powers $x^{7/2}$. We verified that we need to go this power 
in the series if energies up to $x=1\times10^{-2}$ are going to be reproduced. 
Regarding the logarithmic terms,  we did not observe any improvement in the law 
(\ref{eqnostra}) with their inclusion and thus, we finally removed them. The 
best set of parameters are reported in Table \ref{tablenosaltres}, with a 
value $\chi^2/\nu < 1$. The equations of state for different $r_0$ values  in the range $|r_0| < 2$   are
shown, together with the DMC energies, in Fig. \ref{eos_dmc}. As one can 
appreciate, Eq. (\ref{eqnostra}) is able to reproduce quite accurately the 
exact energies. For dilute Fermi gases, it is well known that series beyond the 
universal terms include both the $s$-wave effective range and the 
$p$-wave scattering length $a_1$ \cite{bishop,pera2024}.
In the range of $r_0$ values used for the fit of Eq. (\ref{eqnostra}), we tried to
include also the respective values of $a_1$ that we also known (see Ref.~\cite{suplement}).
We used different models to include the $a_1$ values in the equation of state. However,
we were not able to discern any effect by introducing these values in the fit. This can also be understood
by looking at our results plotted in Fig.~\ref{eos_dmc}. There, we show results that have in common $r_0$ but not $a_1$
and any meaningful difference is not appreciated.

\begin{figure}[b]
    \centering
    \includegraphics[width=0.9\linewidth]{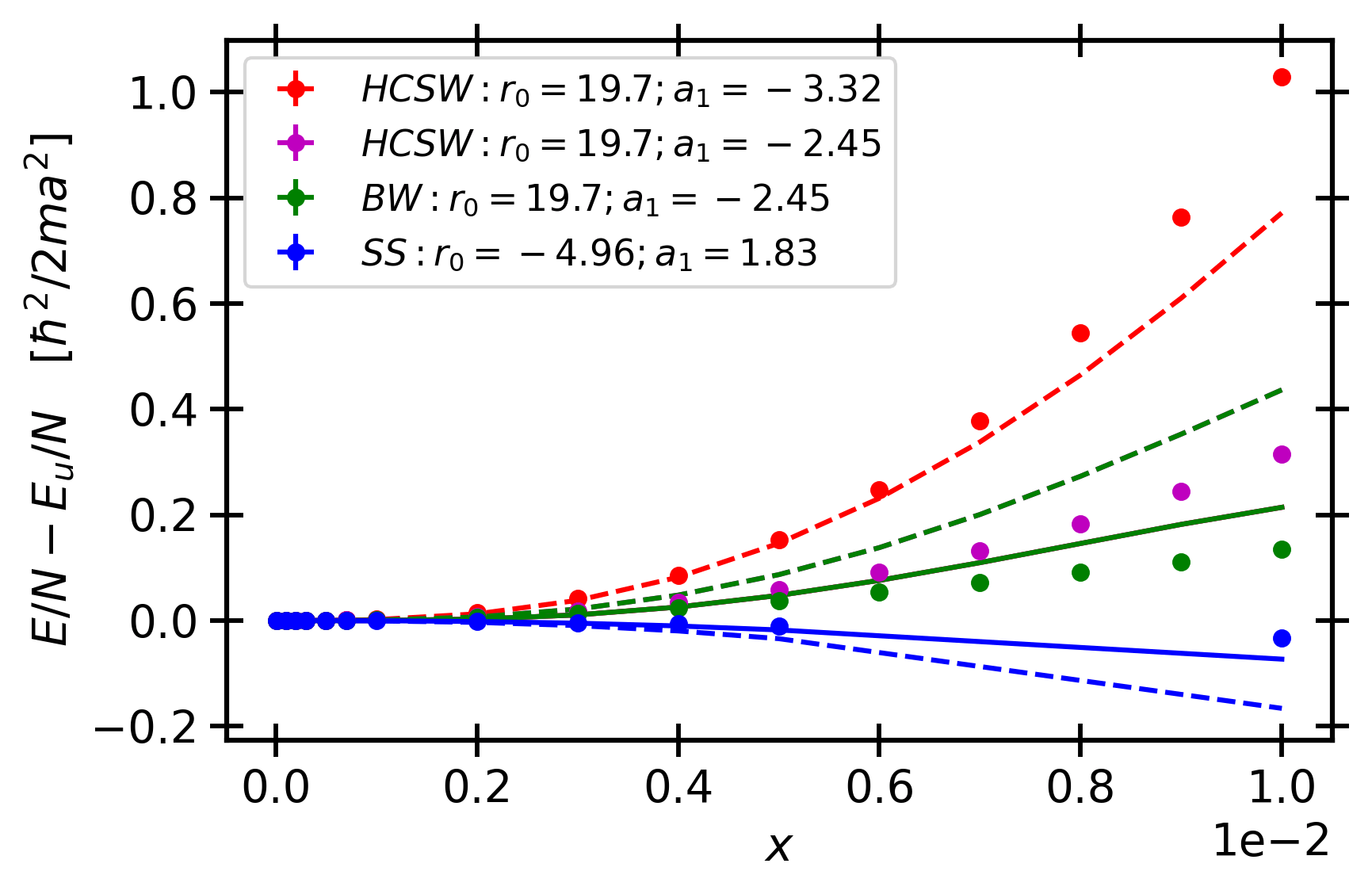}
      \caption{DMC energies for large $|r_0|$ not included in the fit of Eq. 
(\ref{eqnostra}). The error bars are smaller than the size of the symbols. The 
solid lines stand for the application of Eq. (\ref{eqnostra}) and the dashed 
lines incorporate the value of $a_1$ in the equation of state, as commented in 
the text. }
    \label{r0gran}
\end{figure}

We carried out more DMC calculations covering larger values of the effective range.
However, we were not able to get a reliable equation of state, like Eq. (\ref{eqnostra}),
including this extended regime. In Fig.~\ref{r0gran}, we report the energies obtained for these
additional cases, with $r_0=19.7$ and $r_0=-4.96$. For the positive $r_0$ case, 
we use two different potentials,
HCSW and BW, and the negative case corresponds to a SS model. The first 
observation on these results is the breaking of universality, in terms of the 
pair $(a,r_0)$, for the case of large $r_0$ at gas parameters $x \gtrsim 
3\times 10^{-3}$. Therefore, the two-parameters universal range depends on 
the $|r_0|$ value, becoming smaller when it grows. The case of the large $r_0$ 
is particularly interesting. In the figure, we show three different 
calculations 
with a common $r_0$ value, two of them sharing the same $p$-wave scattering 
length. Even being it not perfect, we can see that the energies when the 
triplet $(a,r_0,a_1)$ is the same are closer. This indicates that the role of 
$a_1$, which was negligible for small $|r_0|$, becomes significant when the 
effective range is larger.     

We have analyzed if the equation of state (\ref{eqnostra}), fitted to a 
regime of small effective ranges, is able to reproduce cases out of this 
domain, even in an approximate form. In Fig.~\ref{r0gran}, we show with solid 
lines the predictions of Eq. (\ref{eqnostra}) with  the parameters contained in 
Table \ref{tablenosaltres} for the two new cases with  $r_0=19.7$ and 
$r_0=-4.96$. The non-universal corrections in both cases are significantly 
larger than the cases analyzed before (see the different energy scales in Figs. 
\ref{energies_dmc} and \ref{r0gran}) but still the predictions are rather close 
to the exact results. Of course, Eq.(\ref{eqnostra}) can not reproduce the 
dependence on $a_1$ observed for $r_0=19.7$ because this scattering parameter 
is not included in the model. For the Bose equation of state, there is not any 
perturbative model which includes $a_1$, to the best of our knowledge. 
Noticeably, in the case of fermions $a_1$ appears explicitly in the expansion, 
combined linearly with the effective range. As a first and rather crude 
attempt, we summed $a_1^3$ to $r_0$  in the term $\bar{x}^{3/2}$ of Eq. 
(\ref{eqnostra}) without any refitting procedure (in a second model we assumed 
$r_0 + \alpha a_1^3$ with $\alpha$ an adjusting parameter, but the 
behavior of the energy with $a_1$ was worse 
than with a fixed $\alpha=1$ value). Notice that we add $a_1^3$ 
because the phase shift of $p$-wave scattering is 
proportional to $k^3 A_1$, $A_1$ being a volume that is usually written as 
$a_1^3$ \cite{pera}. The results obtained are 
shown in Fig. \ref{r0gran} with dashed lines. For the two small $|a_1|$ values, 
the introduction of $a_1$ into the equation of state does not improve the 
results. However, for the largest $|a_1|$ the result is surprisingly accurate. 
It 
is also remarkable that the introduction of this scattering parameter reproduces 
well the ordering of the DMC energies when $a_1$ changes.

Summarizing, we  have explored the equation of state of dilute Bose gases by a 
bunch of DMC calculations of energies, covering a large range of scattering 
parameters and using different model potentials. The goal of this intensive 
work was to provide a valuable set of results in order to explore the behavior 
of 
the equation of state when universality in terms of the $s$-wave scattering 
length ceases. In this regard, this is a continuation of work initiated some 
years ago in Ref. \cite{boronat} that allowed to determine the approximated 
limit of the universality. We have carried out our calculations up to gas 
parameters $x=10^{-2}$, quite far of the limit of universality $x < 10^{-3}$. 
Our results show sizeable deviations of the universal terms, with both positive 
and negative departures. In the regime $|r_0| < 2$, we clearly observe 
universality in terms of $(a,r_0)$, since energies calculated with different 
model potentials, but sharing these two scattering parameters, are 
statistically 
compatible up to $x \simeq 10^{-2}$. In this regime, the equation of state 
proposed by Braaten \cite{braaten} reproduces  our results but in a smaller 
density domain, $ 
x < 4 \times 10^{-3}$. We proposed a new equation of state, with terms up to 
$x^{7/3}$, that reproduces quite accurately the DMC energies up to $x = 10^{-2}$.
We used additional energies corresponding to larger values of $r_0$ to check 
the validity of our law out of the regime used for the fitting. Contrarily to 
the energies for small effective range, we observed a sizable effect of the 
$p$-wave scattering length. By fixing the set of three parameters 
$(a,r_0,a_1)$, 
we recovered universality, although for not so large values of $x$. By 
introducing in the equation of state the $a_1$ length, we improve our model, 
mainly for larger values of $a_1$. We hope that our DMC calculations 
\cite{suplement} can serve 
to stimulate further theoretical work to improve the equation of state of Bose 
gases beyond the universal limit. Also, from the experimental side, the 
estimation of $r_0$ can serve to effectively improve the study of Bose gases in 
regimes where the interactions are stronger. Finally, the introduction of the 
new equation of state in the extended Gross-Pitaevskii formalism 
\cite{giorgini_rmp} can enlarge 
significantly its domain of applicability.

We acknowledge Joaquim Casulleras for useful discussions to solve the 
inverse scattering problem. 
We acknowledge financial support from Ministerio de Ciencia e Innovaci\'on
MCIN/AEI/10.13039/501100011033
(Spain) under Grant No. PID2020-113565GB-C21 and
from AGAUR-Generalitat de Catalunya Grant No. 2021-SGR-01411.


\bibliography{bibliography}

\pagebreak
\widetext 
\begin{center}
	\textbf{{\Large Supplemental Material for ``Equation of state of Bose gases beyond the 
universal regime''
	}}
\end{center}






 \renewcommand{\theequation}{S\arabic{equation}}
 \setcounter{equation}{0}
 \renewcommand{\thefigure}{S\arabic{figure}}
 \setcounter{figure}{0}
 \renewcommand{\thesection}{S\arabic{section}}
 \setcounter{section}{0}
 \onecolumngrid
 




\section{Trial wave functions}
\label{Sec:trialf}


The  DMC algorithm  uses a trial or guiding wave function for importance 
sampling. We use a Jastrow model,  
\begin{equation}
    \Psi_T(\mathbf{R})=F_J(\mathbf{R})\Phi(\mathbf{R}) \ . 
\label{trialfjastrow}
\end{equation}
The  one-body term $\Phi(\mathbf{R})$, takes care of the symmetry or 
antisymmetry of the trial wave function, that is, if particles are bosons or 
fermions, respectively, as well as other global properties of the system. 
For a bulk Bose system one just takes $\Phi(\mathbf{R})=1$. The second term 
is the product of two-body correlation functions $f(r_{ij})$ for all pairs $i,j 
= 1,...,N$ in the system,
\begin{equation}
    F_J(\mathbf{R})=\prod_{i<j}f(r_{ij}) \ . \label{jastrow}
\end{equation}
For a central potential, the two-body factor $f(r_{ij})$ is solely 
determined by the distance between the pair of particles. In this work,
we use as a Jastrow factor the exact solution of the two-body problem. In the 
following, we discuss the explicit forms of $f(r)$ for the different model 
potentials used in our study.


As our simulations are intended for a bulk system, that we model using 
periodic boundary conditions, we need to specify the function $f(r)$ for any 
interparticle distance $r\leq L/2$. We tried two different options: i) match 
the two-body solution with a smooth function (exponential) going to a constant 
at the end of that distance, and (ii) to use in all the domain the two-body 
solution, properly symmetrized. We used both methods and obtained 
fully compatible results. In the following, we describe the second method, 
which is easier to build.

We impose that the two-body correlation  function $f(r)$ satisfies 
\begin{align}
   f(L/2)=1, \qquad
    f'(L/2)=0.
\end{align}
For practical reasons, the wave function $\mathcal{R}(r)$ needs to be 
understood as an exponential, that means $\mathcal{R}(r)=e^{v(r)}$. Its 
derivative, in terms of $v(r)$ is $\mathcal{R}'(r)=v'(r)e^{v(r)}$ Then, the 
previous conditions for $v(r)$, but already for the two-body correlation 
$f(r)=e^{\hat{v}(r)}$ result in
\begin{align}
    e^{\hat{v}(L/2)}&=1 \rightarrow \hat{v}(L/2)=0, \label{cond1} \\
    \hat{v}'(L/2)e^{\hat{v}(L/2)}&=0 \rightarrow  \hat{v}'(L/2)=0  .\label{cond2} 
\end{align}
A solution for these two conditions is
\begin{equation}
    \hat{v}(r)=v(r)+v(L-r)-2v(L/2). \label{symm}
\end{equation}
Actually, this is a general transformation for any function $v(r)$ 
to a function $\hat{v}(r)$ that is symmetric with respect to $r=L/2$ in the 
dominion $0<r<L$. Its first and second derivatives are
\begin{align}
        \hat{v}'(r)&=v'(r)-v'(L-r),\label{vder1} \\
        \hat{v}''(r)&=v''(r)+v''(L-r).\label{vder2}
\end{align}
One may verify that, apart from the symmetry, the latter satisfies \eqref{cond1} and \eqref{cond2}:
\begin{align}
        \hat{v}(L/2)&=v(L/2)+v(L-L/2)-2v(L/2)=0 ,\\
        \hat{v}'(L/2)&=v'(L/2)-v'(L-L/2)=0.
\end{align}
In the following, we report the detailed solution for 
each tested potential.

\paragraph{Hard-core square-well (HCSW)}

To find the solution $f(r)$, the starting point is the two-body solution ${\mathcal{R}(r)}$. In the case of the HCSW interaction potential, that is 
\begin{equation}
    \mathcal{R}(r) = \begin{cases} 0    &    0 \leq  r<R_1    \\
    \frac{A}{r}\sin[k_1(r-R_1)]    & R_1 \leq r < R_2 \\
                   \frac{B}{r}\sin[k_2(r-\delta)]    &     r \geq R_2
    \end{cases}
\end{equation}
 In terms of $v(r)=\ln{\mathcal{R}(r)}$,
\begin{equation}
    v(r) = \begin{cases} -\infty    &    0 \leq  r<R_1    \\
    \ln{\frac{A}{r}\sin[k_1(r-R_1)]}    & R_1 \leq r < R_2 \\
                   \ln{\frac{B}{r}\sin[k_2(r-\delta)]}    &     r \geq R_2
    \end{cases}
\end{equation}

The parameters $A$, $B$, $\delta$ and $k_2$ are such that verify the boundary 
conditions of both the function and its derivative.
 The resulting equations are solved for 
$\delta$ and $k_2$, and $A$, $B$ are left as variational parameters. In fact, 
only its relation in the form of the quotient $A/B$ is left, as will be seen, 
since $A$ and $B$ are amplitudes and their absolute value does not affect that 
$f(L/2)=1$. One gets 
\begin{align}
   A \sin [k_1 (R_2-R_1)] &= B \sin[k_2 (R_2 - \delta)]  \label{eqhcsw1}\\
    A k_1 \cos[k_1 (R_2 -R_1)] &= B k_2 \cos[k_2 (R_2 - \delta)] \label{eqhcsw2}
\end{align}
The resolution follows isolating $\delta$ from \eqref{eqhcsw1},
\begin{equation}
    \delta=R_2 - \frac{1}{k_2}\displaystyle\left \{ \arcsin{\displaystyle\left (\frac{A}{B}\sin [k_1 (R_2-R_1)]  \right )} \right\} .\label{deltahcswt2}
\end{equation}
Then, dividing \eqref{eqhcsw1}/\eqref{eqhcsw1}, the transcendental equation for $k_2$ is obtained
\begin{equation}
    \frac{\tan[k_1(R_2-R_1)]}{k_1}=\frac{\tan[k_2(R_2-\delta)]}{k_2},
\end{equation}
since $k_1(k_2)=\sqrt{V_1 + k_2^2}$ and $\delta=\delta(k_2)$ in \eqref{deltahcswt2} (where the dependence on $A/B$ emerges). Defining $u=k_2$, the latter is transformed into the zero of the function
\begin{equation}
    f(u)= \frac{\tan[k_1(u)(R_2-R_1)]}{k_1(u)}-\frac{\tan[u(R_2-\delta(u))]}{u} ,\label{trhcsw}
\end{equation}
which can be calculated using the Newton-Raphson method, using that its derivative is
\begin{align*}
f'(u)=& \frac{k_1'(u)(R_2-R_1)}{k_1(u)\cos^2[k_1(u)(R_2-R_1)]}-\frac{k_1'(u)\tan[k_1(u)(R_2-R_1)]}{k_1(u)^2}-\\
&-\frac{R_2-\delta(u)-u\delta'(u)}{u\cos^2[u(R_2-\delta(u))]} + \frac{\tan[u(R_2-\delta(u))]}{u^2}
\end{align*}
where
\begin{align}
    k_1'(u)&=\frac{2u}{2\sqrt{V_1+u^2}}=\frac{u}{k_1(u)},\\
    \delta'(u)&=\frac{\arcsin{\displaystyle\left (\frac{A}{B}\sin [k_1(u) (R_2-R_1)]  \right )} }{u^2} - \frac{\frac{A}{B} \cos[k_1(u)(R_2-R_1)] k_1'(u)(R_2-R_1) }{u\sqrt{1-\displaystyle \left ( \frac{A}{B}\sin [k_1(u) (R_2-R_1)] \right )^2}}.
\end{align}

Once solved for $u=k_2$, $\delta$ is obtained from \eqref{deltahcswt2}. Thus,
the analytic expressions of $\mathcal{R}(r)$ and $v(r)$ are obtained.

From now on, the notation $v_1(r)$, $v_2(r)$ is employed to refer to the solution $v(r)$ in the region $R_1 \leq r < R_2$ and $r \geq R_2$ respectively. Since $v(r)$ is a piece-wise function, $\hat{v}(r)$ also is. To construct it, \eqref{symm} is applied for each region. 

\begin{equation}
    \hat{v}(r) = \begin{cases} -\infty + v_2(L-r)-2v_2(L/2)=-\infty  &    0 \leq  r<R_1    \\
    \hat{v_1}(r)=v_1(r) + v_2(L-r) -2v_2(L/2) & R_1 \leq r < R_2 \\
    \hat{v_2}(r)=v_2(r) + v_2(L-r) - 2v_2(L/2)       &   R_2 \leq r \leq L/2
    \end{cases}
\end{equation}

Note that the term $v(L-r)$ always corresponds to the expression $v_2(L-r)$, because $L-r>R_2$ for $0\leq r \leq L/2 $ given that $R_2<L/2$. Also, $v(L/2)$ corresponds to $v_2(L/2)$ for the same reason. The explicit expressions for the new wave functions are
\begin{align*}
    \hat{v_1}(r)=&  \ln{\frac{A}{r}\sin[k_1(r-R_1)]} +  \ln{\frac{B}{L-r}\sin[k_2(L-r-\delta)]} - 
     2 \ln{\frac{B}{L/2}\sin[k_2(L/2-\delta)]}\\
    \hat{v_2}(r)=&\ln{\frac{B}{r}\sin[k_2(r-\delta)]} +  \ln{\frac{B}{L-r}\sin[k_2(L-r-\delta)]} - 
     2 \ln{\frac{B}{L/2}\sin[k_2(L/2-\delta)]}
\end{align*}
and $f(r)=e^{\hat{v}(r)}$.

\paragraph{Barrier-well (BW)}
\label{bwsymm}
The two-body solution is 
\begin{equation}
    \mathcal{R}(r) = \begin{cases} \frac{A}{r}\sinh[k_1r]     &    0 \leq  r<R_1    \\
    \frac{B}{r}\sin[k_2(r-\delta)]    & R_1 \leq r < R_2 \\
                   \frac{C}{r}\sin[k_3(r-\xi)]    &      r \geq R_2\\     
    \end{cases}
\end{equation}
where $k_1=\sqrt{V_1-k_3^2}$ and $k_2=\sqrt{V_2+k_3^2}$. 
By imposing the right boundary conditions, one gets
\begin{align}
   A \sinh [k_1 R_1] &= B \sin[k_2 (R_1 - \delta)]  \label{bw1}\\
    A k_1 \cosh[k_1 R_1] &= B k_2 \cos[k_2 (R_2 - \delta)] \label{bw2}\\
   B \sin [k_2 (R_2-\delta)] &= C \sin[k_3 (R_2 - \xi)]  \label{bw3}\\
    B k_2 \cos[k_2 (R_2-\delta)] &= C k_3 \cos[k_3 (R_2 - \xi)] \label{bw4}
\end{align}
This system of equations is solved for $C$, $\delta$, $\xi$, and $k_3$. 
Isolating $\delta$ from \eqref{bw1},
\begin{equation}
        \delta=R_1 - \frac{1}{k_2}\displaystyle\left \{ \arcsin{\displaystyle\left (\frac{A}{B}\sinh [k_1 R_1]  \right )} \right\} \label{deltabwt2}
\end{equation}
and combining \eqref{bw1}/\eqref{bw2} and \eqref{bw3}/\eqref{bw4}, the system is reduced to 
\begin{align}
     \frac{\tanh[k_1R_1]}{k_1}&=\frac{\tan[k_2(R_1-\delta)]}{k_2}\\
      \frac{\tan[k_2(R_2-\delta)]}{k_2}&=\frac{\tan[k_3(R_2-\xi)]}{k_3}
\end{align}
Given its form, setting $B=1$, and leaving $A$ as a variational parameter, conveniently adjusted, the solution for $k_3$, $\xi$ is given by the zero of the multivariate function
\begin{equation}
    f(k_3,\xi)=\displaystyle\left (\frac{\tanh[k_1R_1]}{k_1} - \frac{\tan[k_2(R_1-\delta)]}{k_2} , \frac{\tan[k_2(R_2-\delta)]}{k_2}-\frac{\tan[k_3(R_2-\xi)]}{k_3}\right ). \label{trbw}
\end{equation}
Then, $\delta=\delta(k_3)$ is recovered from \eqref{deltabwt2} and $C$ is 
determined from \eqref{bw3},
\begin{equation}
    C=\frac{B \sin [k_2 (R_2-\delta)]}{\sin[k_3 (R_2 - \xi)] }.
\end{equation}

The resulting transcendental equation is solved by means of a multidimensional Newton-Raphson method. However, this time the derivative $f'(k_3,\xi)$ is approximated numerically, due to the complexity of its analytical expression. To explore its solution, $f(k_3,\xi)$ is understood as two surfaces in the $k_3$-$\xi$ plane, given by its first and second components. Then, the initial guess and $A$ values are set provided that both surfaces intersect with the plane $z=0$, meaning the solution exists. 
From the two-body solution $\mathcal{R}(r)=e^{v(r)}$, one only needs to combine it forming
\begin{equation}
    \hat{v}(r) = \begin{cases} \hat{v_1}(r)= v_1(r) + v_3(L-r)-2v_3(L/2)   &    0 \leq  r<R_1    \\
    \hat{v_2}(r)=v_2(r) + v_3(L-r) -2v_3(L/2) & R_1 \leq r < R_2 \\
    \hat{v_3}(r)=v_3(r) + v_3(L-r) - 2v_3(L/2)    &   R_2 \leq r \leq L/2
    \end{cases}
\end{equation}

Again, the term $v(L-r)$ and $v(L/2)$ corresponds to expression for $v(r)$ that holds in the region $R_2\leq r\leq L/2$, since $L-r>R_2$ for $0\leq r \leq L/2 $ given that $R_2<L/2$. The explicit expressions are

\begin{align*}
    \hat{v_1}(r)=&  \ln{\frac{A}{r}\sin[k_1 r]} +  \ln{\frac{C}{L-r}\sin[k_3(L-r-\xi)]} - 
     2 \ln{\frac{C}{L/2}\sin[k_3(L/2-\xi)]}\\
    \hat{v_2}(r)=&\ln{\frac{B}{r}\sin[k_2(r-\delta)]}+ \ln{\frac{C}{L-r}\sin[k_3(L-r-\xi)]} - 
     2 \ln{\frac{C}{L/2}\sin[k_3(L/2-\xi)]}\\
    \hat{v_3}(r)=&\ln{\frac{C}{r}\sin[k_3(r-\xi)]} +\ln{\frac{C}{L-r}\sin[k_3(L-r-\xi)]} - 
     2 \ln{\frac{C}{L/2}\sin[k_3(L/2-\xi)]}
\end{align*}

\paragraph{Soft sphere (SS)}

The two-body solution is readily obtained,
\begin{equation}
    \mathcal{R}(r) = \begin{cases} \frac{A}{r} \sinh[k_1 r]    &    0 \leq  r<R_1    \\
                   \frac{B}{r}\sin[k_2(r-\delta)]    &    r > R_1\\
    \end{cases}
\end{equation}
where $k_2=\sqrt{mE}/\hbar$ and $k_1=\sqrt{mV_1/\hbar^2 - k_2^2}$. Imposing 
boundary conditions,
\begin{align}
   A \sinh [k_1 R_1] &= B \sin[k_2 (R_1 - \delta)]  \label{eqss1}\\
    A k_1 \cosh[k_1 R_1] &= B k_2 \cos[k_2 (R_1 - \delta)] \label{eqss2}
\end{align}

From \eqref{eqss1}, $\delta$ can be isolated
\begin{equation}
    \delta=R_1 - \frac{1}{k_2}\displaystyle\left \{ \arcsin{\displaystyle\left (\frac{A}{B}\sinh [k_1 R_1]  \right )} \right\} \label{deltasst2}.
\end{equation}

Then, as in the previous cases, the solution only depends on the ratio $A/B$, which is left as a variational parameter that will be properly adjusted. Dividing \eqref{eqss1}/\eqref{eqss2}, the transcendental equation for $k_2$ is obtained,
\begin{equation}
    \frac{\tanh[k_1 R_1]}{k_1}=\frac{\tan[k_2(R_1-\delta)]}{k_2}.
\end{equation}

Then, defining $u=k_2$, it becomes the zero of the function
\begin{equation}
    f(u)= \frac{\tanh[k_1(u) R_1]}{k_1(u)}-\frac{\tan[u(R_1-\delta(u))]}{u},
\end{equation}
which can be calculated using the Newton-Raphson method, knowing that its 
derivative is
\begin{align*}
f'(u)=& \frac{k_1'(u)R_1}{k_1(u)\cosh^2[k_1(u)R_1]}-\frac{k_1'(u)\tanh[k_1(u)R_1]}{k_1(u)^2}-\\
&-\frac{R_1-\delta(u)-u\delta'(u)}{u\cos^2[u(R_1-\delta(u))]} + \frac{\tan[u (R_1-\delta(u))]}{u^2}
\end{align*}
where
\begin{align}
    k_1'(u)&=\frac{-2u}{2\sqrt{mV_1/\hbar^2-u^2}}=-\frac{u}{k_1(u)}\\
    \delta'(u)&=\frac{\arcsin{\displaystyle\left (\frac{A}{B}\sinh [k_1(u) R_1]  \right )} }{u^2} - \frac{\frac{A}{B} \cosh[k_1(u) R_1] k_1'(u) R_1 }{u\sqrt{1-\displaystyle \left ( \frac{A}{B}\sinh [k_1(u) R_1] \right )^2}}
\end{align}

Once solved for $u=k_2$, $\delta$ is obtained from \eqref{deltasst2}. Thus, the
analytic expressions of $\mathcal{R}(r)$ and $v(r)$ are known.
Then, the expressions for $\hat{v}(r)$ are
\begin{equation}
    \hat{v}(r) = \begin{cases} \hat{v_1}(r)=v_1(r) + v_2(L-r)-2v_2(L/2) &    0 \leq  r < R_1    \\
    \hat{v_2}(r)=v_2(r) + v_2(L-r) -2v_2(L/2) &  r\geq R_1 \\
    \end{cases}
\end{equation}

following the same reasoning as with the previous potentials.

\section{Solution of the inverse problem }\label{Sec:scatt}

Given an interaction potential $V(r)$, we are interested in its scattering 
parameters $r_0$ ($s$-wave effective range), $a_0$ (s-wave scattering length) 
and $a_1$ (p-wave scattering length). These are given by
\begin{align}
    a_l^{2l+1} &= \frac{1}{2l+1}\int_0^\infty V(r)r^{l+1}u_l^{(0)}(r)dr, \label{a}\\
    r_0^{eff}&=\frac{2}{a_0^2}\int_0^\infty \displaystyle \left [ (r-a_0)^2 - {u_0^{(0)}}^2(r)\right ]dr, \label{r}
\end{align}
where $l$ is the angular momentum quantum number and has been already 
substituted by $l=0$ 
in Eq.~\eqref{r}. The function $u_l^{(0)}(r)$ is the reduced radial wave 
function, solution to the radial two-body Schr\"odinger equation for that 
potential, with zero energy and angular momentum $l$. It must verify the 
boundary condition $u_l^{(0)}(r=0)=0$ (the wave function must vanish at the 
origin) and be properly normalized to behave as $r^{l+1}-a_l^{2l+1}r^{-l}$ when 
$r\rightarrow \infty$.  

 Our interest lies in comparing the energy of systems described by  
different 
 interaction potentials, each of them with its scattering parameters. To 
do so, one needs to be able to calculate the scattering parameters of the 
interaction potentials employed. We refer to Ref.~\cite{pera} for their 
explicit analytical expressions, which result from solving 
Eqs.~(\eqref{a},\eqref{r}) for each potential. All the interaction potential 
models employed are simple, piece-wise functions and can be tuned through a 
small number of parameters. As an example, if the potential model presents a 
finite barrier, one can obtain different potentials by changing its width 
(distance) and height (potential). The referred equations explicitly relate the 
model's parameters with the scattering parameters. Thus, theoretically one can 
fix $n$ scattering parameters for a potential model that presents $n$ tuning 
options. This is exactly what we are interested in: to select any value of the 
scattering parameters and find the exact different potentials that lead to them, 
to study the dependencies at wish.  The problem here lies in the fact that this 
association leads to a non-trivial system of equations.

As an example, consider the barrier-well potential. This model consists 
of a finite barrier  and a well around it. 
The number of tuning options is four: barrier length $R_1$, well length $R_2$, 
barrier height $V_1$ and well height $V_2$. Then, one could solve a system of 
equations for up to four scattering parameters. Since we are only interested in 
three of them, the system of equations is given by: $a_0(R_1,R_2,V_1,V_2)$, 
$r_0(R_1,R_2,V_1,V_2)$, $a_1(R_1,R_2,V_1,V_2)$. The expressions can be found in 
\cite{pera}, by switching $k_1$, $k_2$ to $ik_1$, $ik_2$ respectively from the 
well-barrier equations. This can be done because the barrier-well model is the 
inverse of the well-barrier, meaning the only difference is a change of sign of 
the potential values, which affects the equations by turning $k_1$ and $k_2$ 
(the square root of the potential values) imaginary. Then, one could attempt to 
fix a triad $a_0$, $r_0$, $a_1$ and $R_1$ and find the values of $R_2$, $V_1$ 
and $V_2$ that verify the system of equations. It is not possible to isolate 
analytically the unknowns from the system of equations, so a numerical method is 
the only way to proceed. A usual approach is the multidimensional Newton-Raphson 
method. However, only that does not solve the problem in general; previously it 
is necessary that one: (i) ensures that there exists a solution for the set of 
parameters fixed and (ii) provides an initial guess for the unknown variables 
that is close to the solution of the system of equations.

We use a stochastic strategy to explore the scattering equations, 
by solving the inverse problem. That means, trying different combinations 
$\{R_1,R_2,V_1,V_2\}$ and calculating the scattering parameters for each 
set, employing the equations provided. Then, each combination leads to a 
point in the $(a_0,a_1,r_0)$ space. Repeating this process for random 
combinations (all of them feasible for that potential model) provides an idea of 
the images of the scattering equations. In the stochastic sense, each 
combination is a choice and provides a triad which is solution of the scattering 
parameters. Then, the regions with more density of points are those where it is 
more likely to find a triad solution. Conversely, the regions with no points are 
not likely reporting a potential based on that model with those scattering 
parameters. By inspecting the whole picture, one has an idea about what 
scattering parameters' triads can be obtained, and, through a simple algorithm, 
recovers the parameters of the potential model that lead to that triad. Finally, 
those potential parameters serve as the initial condition for the Newton-Raphson 
method to find triads similar to the one that is obtained by them, up to the 
desired accuracy.

\begin{figure}[ht]
    \centering
    \subfigure[ Zoom out.]{
        \includegraphics[width=0.45\textwidth]{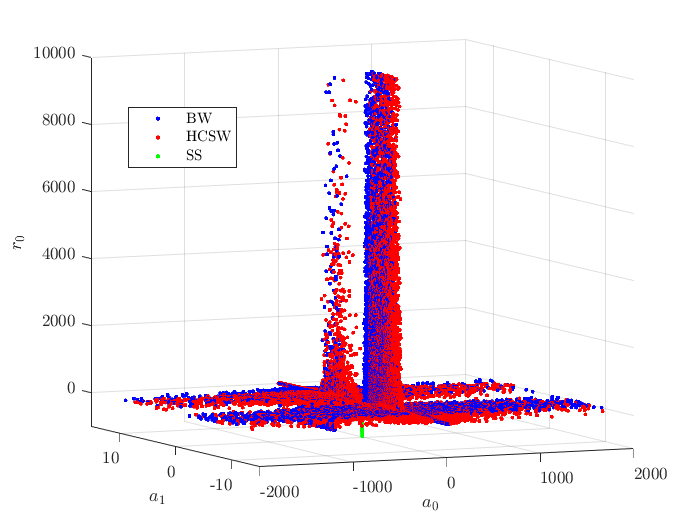}
        }
    \hfill
    \subfigure[ Zoom in.]{
        \includegraphics[width=0.45\textwidth]{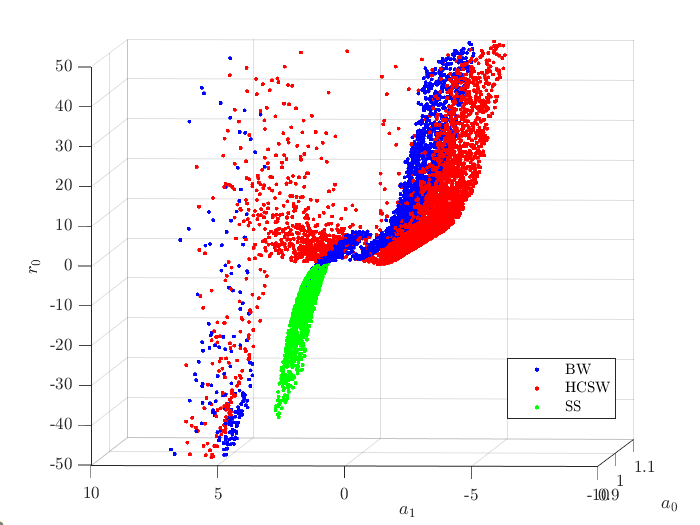}
        }
    \caption{Space of solutions $(a_0,a_1,r_0)$ of the scattering equations for different potentials (see legend) for $10^6$ points.}
    \label{fig:scatparams}
\end{figure}

This strategy has an even major interest in our scope: it allows us to 
compare the images of the scattering parameters' equations of different 
potential models (Fig.~\ref{fig:scatparams}). Thus, it can be extremely useful 
to find coincident triads for different potential models and, if possible, which 
are approximately its values. This case corresponds to a region 
where two points from different potentials are close in distance.    

\begin{figure}[ht]
    \centering
    \subfigure[ Zoom out.]{
        \includegraphics[width=0.45\textwidth]{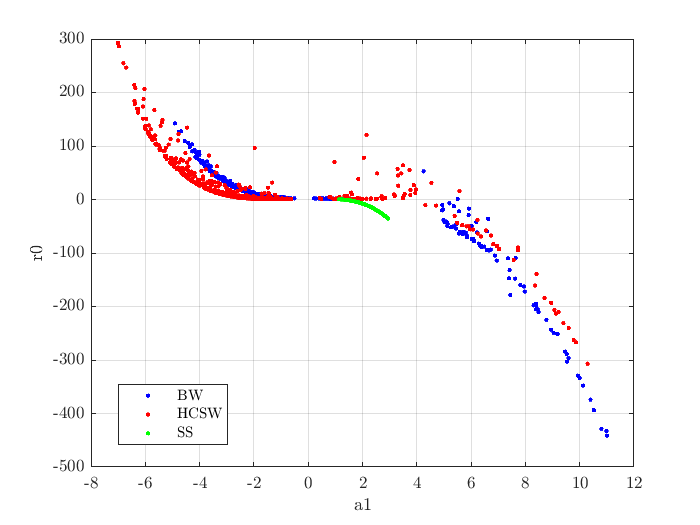}
        }
    \hfill
    \subfigure[ Zoom in.]{
        \includegraphics[width=0.45\textwidth]{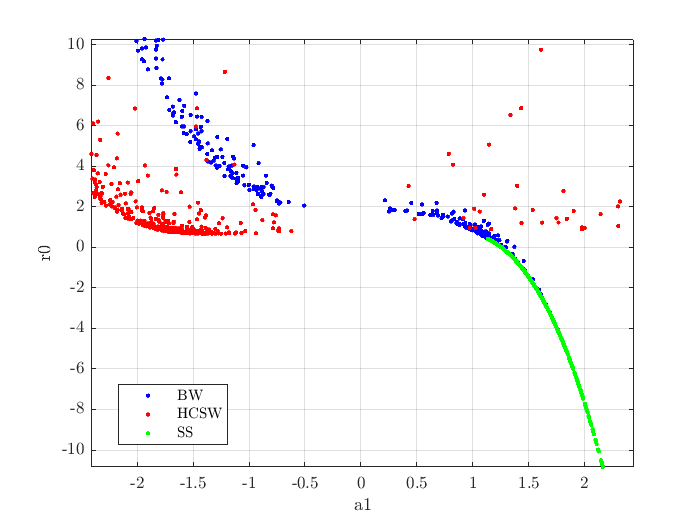}
        }
    \caption{Slice $a_0\approx 1$ in the space of solutions of the scattering equations for different potentials (see legend) for $10^8$ points.}
    \label{fig:scatparams_a0}
\end{figure}

We choose $a_0=1$ for all the interaction potentials, such that 
the distance is in units of $a_0$ in the DMC calculations. Hence, we are 
interested in the points contained in the plane $a_0=1$. In fact,  
Fig.~\ref{fig:scatparams_a0} shows the points contained in 
$a_0\in[1-\delta,1+\delta]$. Recall that due to the stochastic nature of this 
strategy, it is not possible to obtain any point with $a_0=1$ exactly, thus we 
select those that are sufficiently close such that they serve as an initial 
condition for the Newton-Raphson method to find a solution with $a_0=1$ exactly.

Once a solution triad is found, with its associated potential model parameters, 
we use them to calculate the scattering parameters differently again (solving 
the integral forms \eqref{a}, \eqref{r}), such that if they coincide the 
solution is validated. Also, it is checked that a bound state does not appear, 
the $a_0$ value of the triad solution being always at the left of the first 
Feshbach resonance. 

\section{Correction of finite-size effects in DMC}
 
A DMC computation is set for a specific number of particles $N$ at a given 
density $n$, defining a cubic box of side $L=\sqrt[3]{N/n}$. Obviously, $N$ is 
limited by the increase of computer time, that grows as $\sim N^2$. We need to 
estimate finite-size effects because our calculations are intended for a bulk 
phase in the thermodynamic limit.

\begin{figure}[htbp]
    \centering
    \subfigure[  Linear fit to the DMC energies in terms of $1/N$.]{%
        \includegraphics[width=0.55\textwidth]{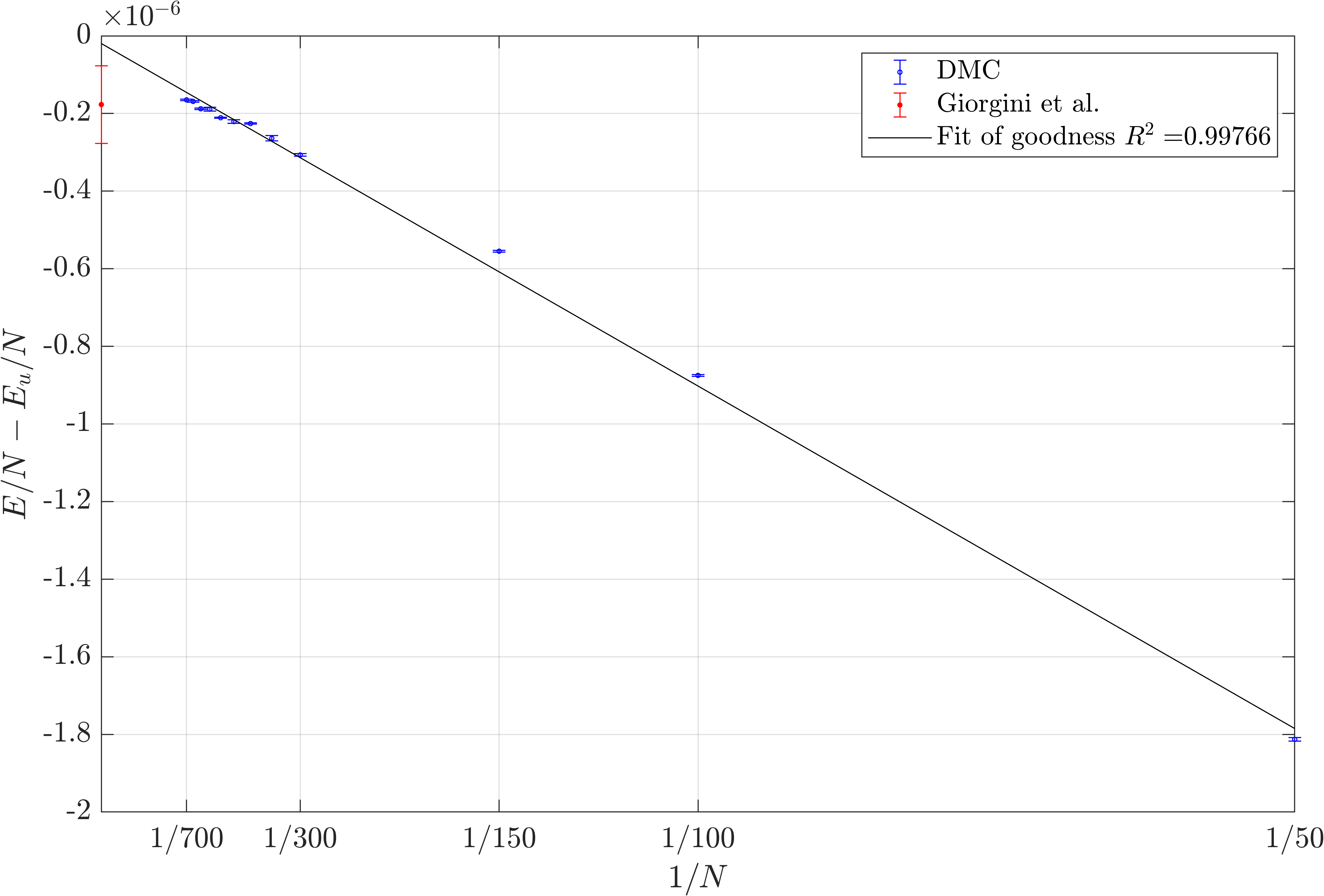}
        \label{fig_1e-5_reg}
    }
    \hfill
    \subfigure[  Extrapolation using data obtained with $N=150$ and 
300. ]{%
        \includegraphics[width=0.43\textwidth]{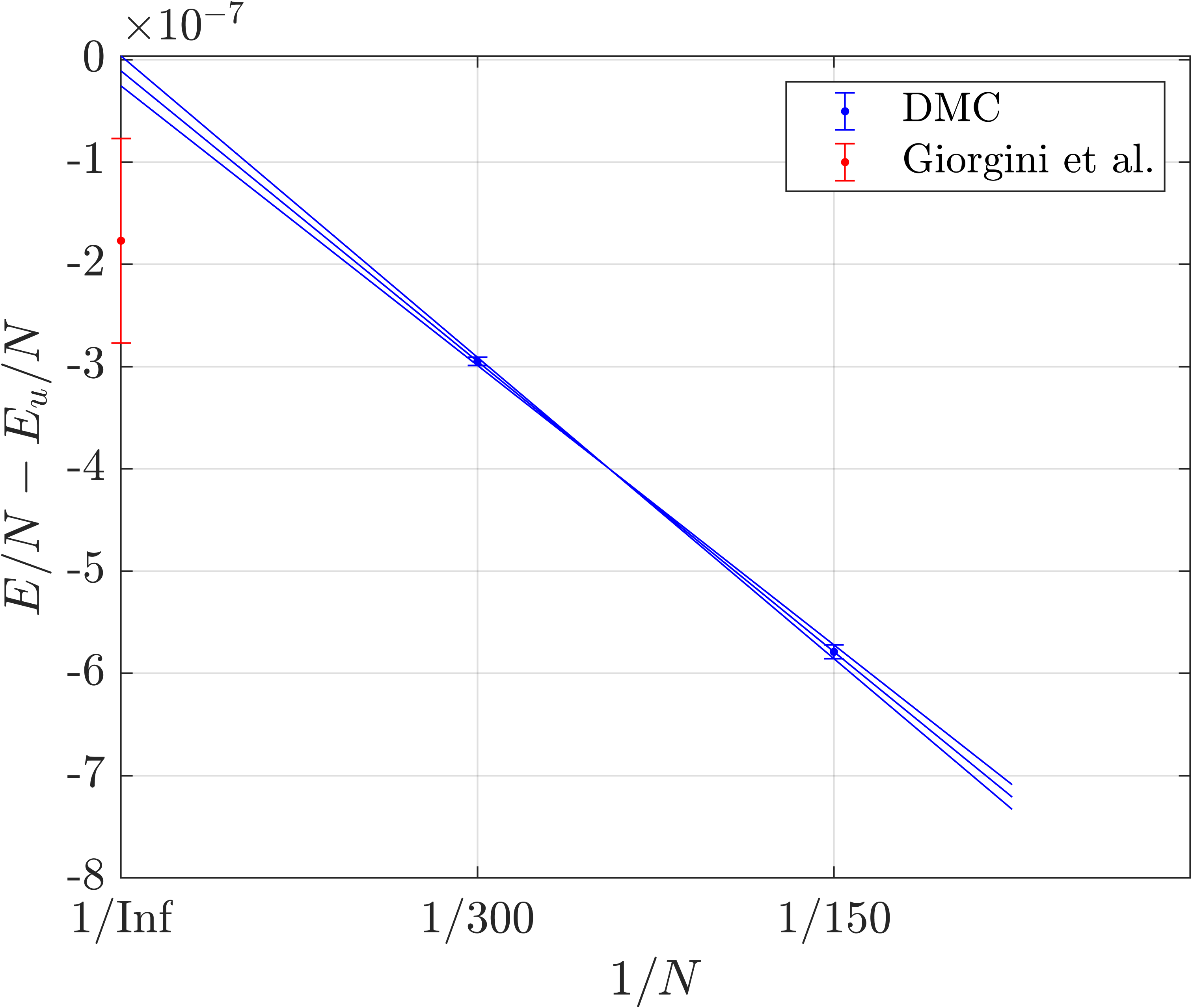}
        \label{fig_1e-5_extrap}
    }
    \caption{DMC energies for different numbers of particles interacting 
through a Hard-Sphere potential at density $na_0^3=10^{-5}$. }
    \label{fig_1e-5}
\end{figure}

\begin{figure}[htbp]
    \centering
    \subfigure[  Linear fit  to the DMC energies in terms of $1/N$.]{%
        \includegraphics[width=0.55\textwidth]{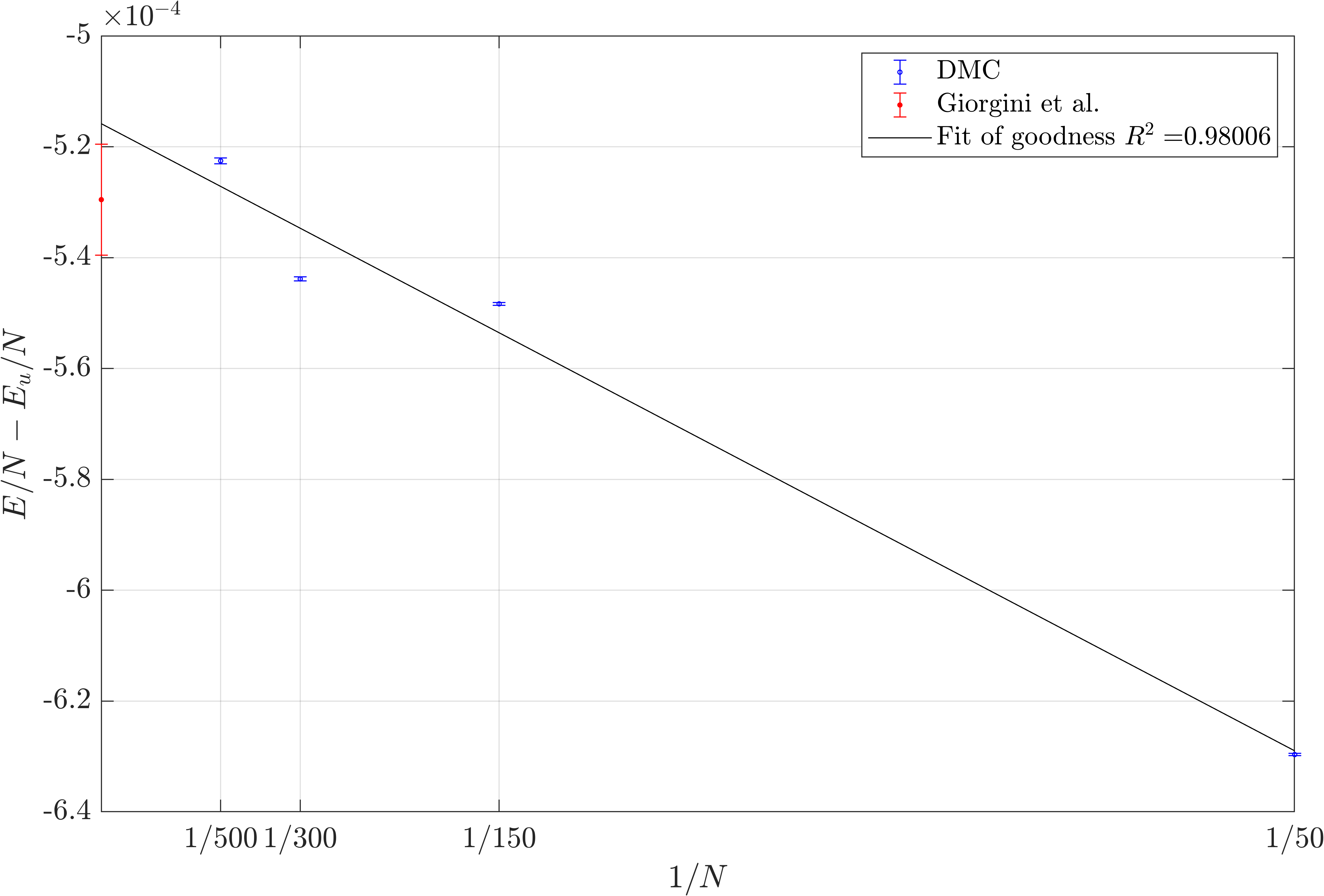}
        \label{fig_1e-3_reg}
    }
    \hfill
    \subfigure[Extrapolation using data obtained with $N=150$ and 
300. ]{%
        \includegraphics[width=0.43\textwidth]{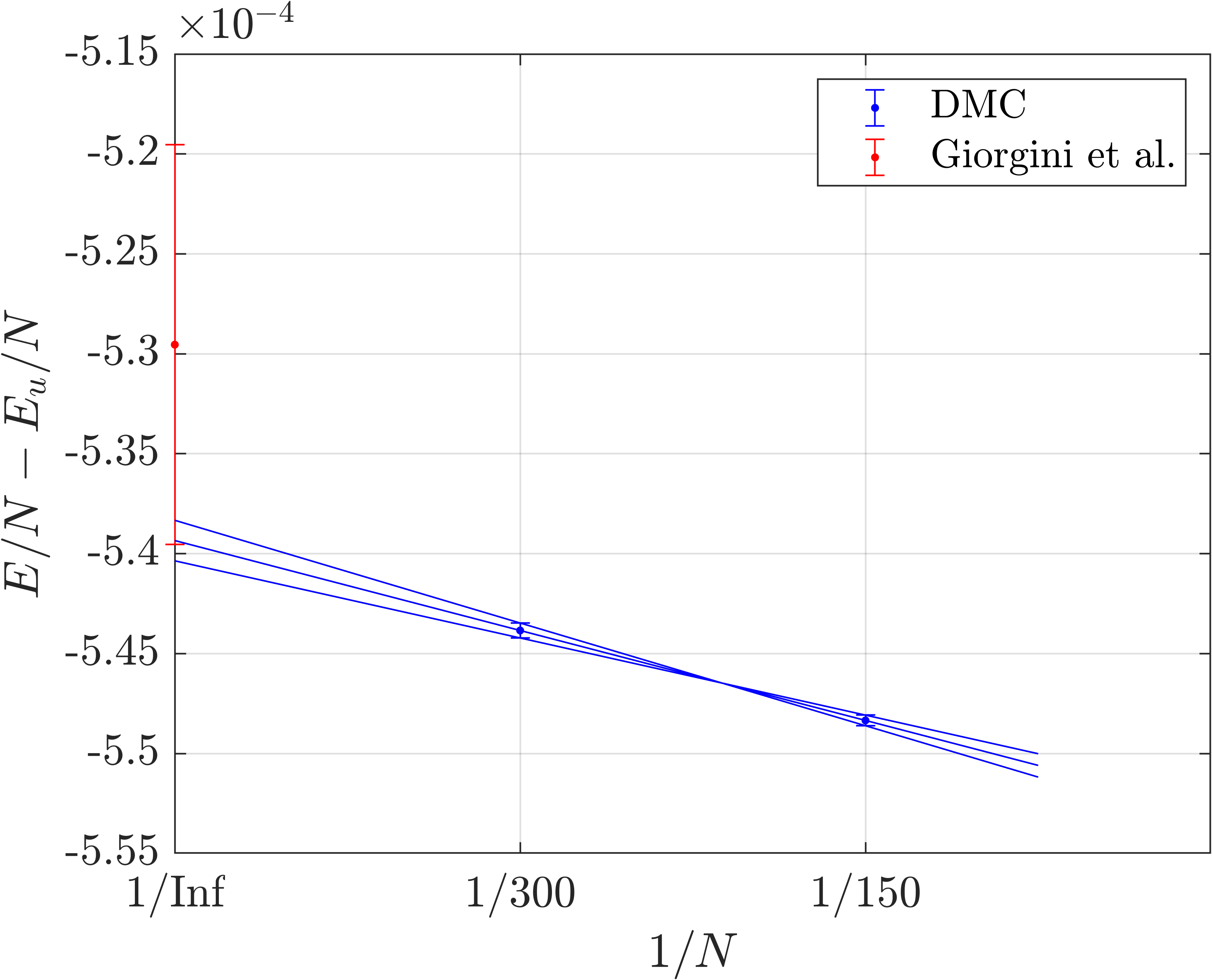}
        \label{fig_1e-3_extrap}
    }
    \caption{DMC energies for different numbers of particles interacting 
through a Soft-Sphere potential (with scattering parameters $r_0=-4.96$, 
$a_1=1.83$) at density $na^3=10^{-3}$. }
    \label{fig_1e-3}
\end{figure}

To overcome this issue, we perform several DMC calculations with different 
numbers of particles and, as a check,  we compare our results with $E_u/N$ at a 
density $na_0^3=10^{-5}$ (Fig.~\ref{fig_1e-5}) where universality holds. A 
similar analysis is shown in Fig.~\ref{fig_1e-3} where $E_u/N$ is no longer 
valid.  Notice that for efficiency reasons in the DMC branching process, 
the product $E\,\Delta t$ needs to remain 
approximately constant, hence the time-step $\Delta t$ should be reduced when 
$N$, and thus $E$, increases. As an 
approximation, we deal with this by setting the product $N \, \Delta t$ 
constant along DMC calculations with different numbers of particles, at each 
density and interatomic potential.

 We observe that the difference between the energy per particle in the DMC 
calculation $E/N$ and the universal energy $E_u/N$
is linear with the inverse number of particles 
$1/N$. At low density, this statement holds in the light of Figure 
{\color{blue}{S}}\ref{fig_1e-5_reg}; while for larger density, 
Fig.~{\color{blue}{S}}\ref{fig_1e-3_reg}  
the linear law is less accurate. As one can see, at low density the agreement 
between the extrapolation to $1/N \to 0$ and the expected result is perfect, 
improving slightly previous DMC estimations on the same system~\cite{boronat}.

Based on the check at low density, all our DMC results consist of calculating 
the energy with $N=150$ and $N=300$, and then extrapolate linearly to 
the limit $1/N \to 0$.  The errors are safely propagated 
accordingly. At low density (Fig.~{\color{blue}{S}}\ref{fig_1e-5_extrap}), this 
procedure leads 
to reproducing $E_u/N$. 

\pagebreak

\section{DMC results}\label{Sec:tables}

\begin{table}[ht]
    \centering
    \begin{tabular}{|c||c|c|c|c|c|c|c|}
        \hline
        Model & $a_0$ & $r_0$ & $a_1$ & $V_1$ & $V_2$ & $R_1$ & $R_2$ \\
        \hline\hline
       SS  & $1$ & $-4.963717$ & $1.828551$ & $3.154281E-2$ & -- & 5 & -- \\
        \hline
        SS & $1$ & $0.246789895603954$ & $1.192367656777861$ & $0.553699098716986$ & -- & $2.25305223642389$ & -- \\
        \hline
       SS & $1$ &$0$ & $ 1.257306682690198$ & $0.352264068783665$ & -- & $2.52455873378522$ & -- \\
        \hline
       HCSW & $1$ & $19.7493$ & $-2.44558$ & $2.816293678E-2$ & -- &$2.23871367134203$ &$6.87815732919500$ \\
        \hline
        HCSW  & $1$ & $19.7493$ & $-3.32211$ & $8.879685093760E-2$& -- & $2.65721897814958$ & $5.91964724378216$ \\
        \hline
       HCSW  & $1$ & $0.951316$ & $-1.6636$& $0.212152394371251$ & -- &  $1.24099358420681$& $2.65518281462878$\\
        \hline
        HCSW & $1$ & $0.951316$ & $ 2.00307$& $7.4019719854E-3$ & -- & $1.07208359841336$ & $4.12341010567992$ \\
        \hline
        HCSW  & $1$ & $0.6842384$ &$1.05598013574085$ & $3.54339300812E-3$& -- & $1.01028825874101$ & $3.06371922192216$ \\
        \hline
        HCSW & $1$ & $0.6842384$ & $-1.4545$ & $7.14004450115E-2$ & -- & $1.03334693697237$ & $2.13914846375276$ \\
        \hline
        HCSW & $1$ & $1.8227872$ & $0.932469981150490$ & $3.82475925030E-2$& -- & $1.29735440984730$ &$4.04020841908368$ \\
        \hline
        HCSW & $1$ & $0.7$ & $0$ & $1.01776107899E-2$ & --& $1.02287211424451$  & $2.90289035827995$ \\
        \hline
        BW & $1$ &$19.7493$ & $-2.44558$ &$0.393919355038301$ &$2.088440731E-2$ & $3.73355462692938$ & $7.312800$ \\
        \hline
        BW & $1$ & $0.6842384$ & $1.05598013574085$ & $0.8005690524$&$0.101545616787338$ & $2.28829062303712 $ & $3.09329071235972$ \\
        \hline
        BW & $1$ & $1.192368$ & $0.246789895603954$ &$0.55130064452$ & $5.7315830258E-3$ & $ 2.26041796351839$ & $2.62247347275118$ \\
        \hline

    \end{tabular}
    \caption{Parameters of the two-body interaction potentials.}
    \label{tab:example}
\end{table}

\begin{table}[h!]
\centering
\begin{tabular}{c|c|c|c|c}

  & HCSW & HCSW & HCSW & HCSW \\
$na_0^3$    &   $r_0=0.95 $    & $r_0=0.95$    &  $r_0=0.68$   & $r_0=0.68$      
 \\
    &   $a_1 = -1.66 $   & $a_1= 2.00$       &  $a_1=1.05$   & $a_1=-1.45$    \\
\hline\hline
$1.0E-5$  & $1.2633E-4 \pm 8.7E-7$ & $1.266E-4 \pm 1.1E-6$ & $1.2695E-4 \pm 6.5E-7$ & $1.2683E-4 \pm 8.5E-7$ \\ \hline
$1.0E-4$  & $1.306E-3 \pm 1.1E-5$ & $1.307E-3 \pm 1.2E-5$ & $1.306E-3 \pm 1.2E-5$ & $1.307E-3 \pm 1.1E-5$ \\ \hline
$2.0E-4$  & $2.667E-3 \pm 1.7E-5$ & $2.658E-3 \pm 2.0E-5$ & $2.658E-3 \pm 2.0E-5$ & $2.654E-3 \pm 1.7E-5$ \\ \hline
$3.0E-4$  & $4.050E-3 \pm 3.2E-5$ & $4.040E-3 \pm 2.4E-5$ & $4.037E-3 \pm 2.6E-5$ & $4.038E-3 \pm 2.6E-5$ \\ \hline
$5.0E-4$  & $6.857E-3 \pm 6.5E-5$ & $6.867E-3 \pm 6.8E-5$ & $6.862E-3 \pm 5.8E-5$ & $6.865E-3 \pm 6.1E-5$ \\ \hline
$7.0E-4$  & $9.773E-3 \pm 7.0E-5$ & $9.764E-3 \pm 7.5E-5$ & $9.756E-3 \pm 6.6E-5$ & $9.758E-3 \pm 6.2E-5$ \\ \hline
$1.0E-3$  & $1.424E-2 \pm 1.4E-4$ & $1.422E-2 \pm 1.2E-4$ & $1.420E-2 \pm 1.2E-4$ & $1.420E-2 \pm 1.2E-4$ \\ \hline
$2.0E-3$  & $2.989E-2 \pm 2.6E-4$ & $2.984E-2 \pm 2.9E-4$ & $2.976E-2 \pm 2.1E-4$ & $2.977E-2 \pm 1.8E-4$ \\ \hline
$3.0E-3$  & $4.650E-2 \pm 3.0E-4$ & $4.641E-2 \pm 3.2E-4$ & $4.623E-2 \pm 4.4E-4$ & $4.622E-2 \pm 3.5E-4$ \\ \hline
$4.0E-3$  & $6.375E-2 \pm 3.5E-4$ & $6.382E-2 \pm 3.3E-4$ & $6.339E-2 \pm 4.5E-4$ & $6.342E-2 \pm 4.6E-4$ \\ \hline
$5.0E-3$  & $8.211E-2 \pm 5.3E-4$ & $8.189E-2 \pm 6.5E-4$ & $8.127E-2 \pm 7.4E-4$ & $8.131E-2 \pm 4.8E-4$ \\ \hline
$6.0E-3$  & $1.0086E-1 \pm 6.1E-4$ & $1.0062E-1 \pm 9.4E-4$ & $9.972E-2 \pm 6.3E-4$ & $9.971E-2 \pm 6.8E-4$ \\ \hline
$7.0E-3$  & $1.2040E-1 \pm 7.1E-4$ & $1.202E-1 \pm 1.1E-3$ & $1.1878E-1 \pm 6.5E-4$ & $1.188E-1 \pm 1.1E-3$ \\ \hline
$8.0E-3$  & $1.407E-1 \pm 1.3E-3$ & $1.4008E-1 \pm 9.0E-4$ & $1.3844E-1 \pm 9.5E-4$ & $1.3837E-1 \pm 7.6E-4$ \\ \hline
$9.0E-3$  & $1.613E-1 \pm 1.3E-3$ & $1.608E-1 \pm 1.2E-3$ & $1.5844E-1 \pm 9.8E-4$ & $1.585E-1 \pm 1.5E-3$ \\ \hline
$1.0E-2$  & $1.815E-1 \pm 1.7E-3$ & $1.813E-1 \pm 1.2E-3$ & $1.790E-1 \pm 1.2E-3$ & $1.790E-1 \pm 1.7E-3$ \\ 
\hline
\end{tabular}
\caption{DMC energies.}
\label{table:1}
\end{table}

\begin{table}[h!]
\centering
\begin{tabular}{c|c|c|c|c}
  & BW & BW & SS & HCSW \\
$na_0^3$    &   $r_0=0.68 $    & $r_0=0.25$    &  $r_0=0.25$   & $r_0=1.82$      
 \\
    &   $a_1 = 1.05 $   & $a_1= 1.19$       &  $a_1=1.19$   &     $a_1=0.93$    \\
\hline\hline
$1.0E-5$  & $1.2671E-4 \pm 9.9E-7$ & $1.2684E-4 \pm 9.6E-7$ & $1.269E-4 \pm 1.1E-6$ & $1.257E-4 \pm 1.2E-6$ \\ \hline
$1.0E-4$  & $1.303E-3 \pm 1.2E-5$ & $1.308E-3 \pm 1.1E-5$ & $1.307E-3 \pm 1.3E-5$ & $1.308E-3 \pm 1.0E-5$ \\ \hline
$2.0E-4$  & $2.649E-3 \pm 1.7E-5$ & $2.656E-3 \pm 2.1E-5$ & $2.657E-3 \pm 2.2E-5$ & $2.660E-3 \pm 1.4E-5$ \\ \hline
$3.0E-4$  & $4.032E-3 \pm 3.1E-5$ & $4.030E-3 \pm 3.5E-5$ & $4.033E-3 \pm 3.7E-5$ & $4.045E-3 \pm 4.0E-5$ \\ \hline
$5.0E-4$  & $6.836E-3 \pm 5.7E-5$ & $6.847E-3 \pm 5.9E-5$ & $6.847E-3 \pm 6.0E-5$ & $6.882E-3 \pm 4.3E-5$ \\ \hline
$7.0E-4$  & $9.716E-3 \pm 5.5E-5$ & $9.731E-3 \pm 6.4E-5$ & $9.728E-3 \pm 6.3E-5$ & $9.802E-3 \pm 5.7E-5$ \\ \hline
$1.0E-3$  & $1.403E-2 \pm 1.4E-4$ & $1.416E-2 \pm 1.3E-5$ & $1.4157E-2 \pm 7.5E-5$ & $1.428E-2 \pm 1.1E-4$ \\ \hline
$2.0E-3$  & $2.985E-2 \pm 1.6E-4$ & $2.961E-2 \pm 1.6E-4$ & $2.964E-2 \pm 2.5E-4$ & $3.014E-2 \pm 2.8E-4$ \\ \hline
$3.0E-3$  & $4.609E-2 \pm 3.3E-4$ & $4.579E-2 \pm 3.6E-4$ & $4.585E-2 \pm 4.1E-4$ & $4.721E-2 \pm 2.9E-4$ \\ \hline
$4.0E-3$  & $6.331E-2 \pm 3.5E-4$ & $6.263E-2 \pm 3.5E-4$ & $6.267E-2 \pm 3.1E-4$ & $6.518E-2 \pm 5.3E-4$ \\ \hline
$5.0E-3$  & $8.122E-2 \pm 7.3E-4$ & $8.005E-2 \pm 4.4E-4$ & $7.985E-2 \pm 5.8E-4$ & $8.413E-2 \pm 7.6E-4$ \\ \hline
$6.0E-3$  & $9.982E-2 \pm 7.7E-4$ & $9.785E-2 \pm 8.2E-4$ & $9.794E-2 \pm 9.7E-4$ & $1.0416E-1 \pm 9.3E-4$ \\ \hline
$7.0E-3$  & $1.1839E-1 \pm 9.0E-4$ & $1.1618E-1 \pm 6.8E-4$ & $1.158E-1 \pm 1.1E-3$ & $1.2484E-1 \pm 7.7E-4$ \\ \hline
$8.0E-3$  & $1.3761E-1 \pm 9.5E-4$ & $1.3465E-1 \pm 1.2E-3$ & $1.347E-1 \pm 1.3E-3$ & $1.4646E-1 \pm 8.8E-4$ \\ \hline
$9.0E-3$  & $1.5730E-1 \pm 8.9E-4$ & $1.5386E-1 \pm 1.5E-3$ & $1.538E-1 \pm 1.4E-3$ & $1.691E-1 \pm 1.5E-3$ \\ \hline
$1.0E-2$  & $1.775E-1 \pm 1.6E-3$ & $1.730E-1 \pm 1.6E-3$ & $1.729E-1 \pm 1.7E-3$ & $1.930E-1 \pm 1.9E-3$ \\ \hline

\end{tabular}
\caption{DMC energies.}
\label{table:2}
\end{table}

\begin{table}[h!]
\centering
\begin{tabular}{c|c|c|c|c}
  & HCSW & SS & HCSW & BW \\
$na_0^3$    &   $r_0=0.7$    & $r_0=0$    &  $r_0=19.7493$   & $r_0=19.7493$     
  \\
    &   $a_1 = 0 $   & $a_1= 1.2573$       &  $a_1=-3.32211$   &     $a_1=-2.44558$    \\
\hline\hline
$1.0E-5$  & $1.268E-4 \pm 1.2E-6$ & $1.269E-4 \pm 1.1E-6$ & $1.3092E-4 \pm 7.2E-7$ & $1.270E-4 \pm 1.1E-6$ \\ \hline
$1.0E-4$  & $1.306E-3 \pm 1.2E-5$ & $1.309E-3 \pm 1.3E-5$ & $1.3244E-3 \pm 8.5E-6$ & $1.3363E-3 \pm 9.2E-6$ \\ \hline
$2.0E-4$  & $2.658E-3 \pm 1.4E-5$ & $2.642E-3 \pm 2.4E-5$ & $2.742E-3 \pm 1.7E-5$ & $2.688E-3 \pm 2.2E-5$ \\ \hline
$3.0E-4$  & $4.038E-3 \pm 3.3E-5$ & $4.034E-3 \pm 2.5E-5$ & $4.290E-3 \pm 3.8E-5$ & $4.154E-3 \pm 2.3E-5$ \\ \hline
$5.0E-4$  & $6.863E-3 \pm 6.2E-5$ & $6.846E-3 \pm 4.0E-5$ & $7.569E-3 \pm 4.5E-5$ & $7.295E-3 \pm 4.7E-5$ \\ \hline
$7.0E-4$  & $9.757E-3 \pm 5.6E-5$ & $9.734E-3 \pm 7.4E-5$ & $1.124E-2 \pm 1.1E-4$ & $1.0536E-2 \pm 7.5E-5$ \\ \hline
$1.0E-3$  & $1.4201E-2 \pm 7.1E-5$ & $1.4149E-2 \pm 9.7E-5$ & $1.756E-2 \pm 1.5E-4$ & $1.5733E-2 \pm 8.7E-5$ \\ \hline
$2.0E-3$  & $2.978E-2 \pm 2.0E-4$ & $2.955E-2 \pm 1.6E-4$ & $4.605E-2 \pm 4.1E-4$ & $3.595E-2 \pm 3.1E-4$ \\ \hline
$3.0E-3$  & $4.625E-2 \pm 3.5E-4$ & $4.567E-2 \pm 2.6E-4$ & $8.922E-2 \pm 6.2E-4$ & $6.083E-2 \pm 6.0E-4$ \\ \hline
$4.0E-3$  & $6.342E-2 \pm 4.7E-4$ & $6.231E-2 \pm 3.9E-4$ & $1.515E-1 \pm 1.3E-3$ & $8.939E-2 \pm 6.0E-4$ \\ \hline
$5.0E-3$  & $8.136E-2 \pm 5.6E-4$ & $7.936E-2 \pm 4.6E-4$ & $2.369E-1 \pm 1.7E-3$ & $1.2179E-1 \pm 7.6E-4$ \\ \hline
$6.0E-3$  & $9.977E-2 \pm 8.0E-4$ & $9.681E-2 \pm 5.9E-4$ & $3.512E-1 \pm 2.3E-3$ & $1.568E-1 \pm 1.5E-3$ \\ \hline
$7.0E-3$  & $1.190E-1 \pm 1.1E-3$ & $1.146E-1 \pm 1.0E-3$ & $5.017E-1 \pm 2.6E-3$ & $1.950E-1 \pm 1.0E-3$ \\ \hline
$8.0E-3$  & $1.3850E-1 \pm 7.4E-4$ & $1.330E-1 \pm 1.2E-3$ & $6.886E-1 \pm 3.8E-3$ & $2.357E-1 \pm 1.7E-3$ \\ \hline
$9.0E-3$  & $1.586E-1 \pm 1.3E-3$ & $1.5155E-1 \pm 8.5E-4$ & $9.284E-1 \pm 6.2E-3$ & $2.763E-1 \pm 1.7E-3$ \\ \hline
$1.0E-2$  & $1.791E-1 \pm 1.3E-3$ & $1.703E-1 \pm 1.4E-3$ & $1.2158E-0 \pm 7.3E-3$ & $3.204E-1 \pm 1.9E-3$ \\ \hline

\end{tabular}
\caption{DMC energies.}
\label{table:3}
\end{table}

\begin{table}[h!]
\centering
\begin{tabular}{c|c|c}

  & HCSW & SS  \\
$na_0^3$    &   $r_0=19.7493$    & $r_0=-4.96$          \\
    &   $a_1 = -2.44558 $   & $a_1= 1.83$          \\
\hline\hline
$1.0E-5$  & $1.2717E-4 \pm 9.2E-7$ & $1.2712E-4 \pm 6.8E-7$\\ \hline
$1.0E-4$  & $1.329E-3 \pm 1.2E-5$ & $1.304E-3\pm 1.1E-5$ \\ \hline
$2.0E-4$  & $2.724E-3 \pm 2.5E-5$ & $2.650E-3\pm 2.4E-5$\\ \hline
$3.0E-4$  & $4.209E-3 \pm 3.8E-5$ & $4.025E-3\pm 3.7E-5 $ \\ \hline
$5.0E-4$  & $7.211E-3 \pm 6.4E-5$ & $6.792E-3\pm 5.4E-5$ \\ \hline
$7.0E-4$  & $1.0579E-2 \pm 9.5E-5$ & $9.622E-3\pm 6.8E-5$ \\ \hline
$1.0E-3$  & $1.590E-2 \pm 1.5E-4$ & $1.3927E-2\pm 7.3E-5$\\ \hline
$2.0E-3$  & $3.742E-2 \pm 2.9E-4$ & $2.855E-2\pm 2.4E-4$\\ \hline
$3.0E-3$  & $6.510E-2 \pm 6.3E-4$ & $4.366E-2\pm 3.1E-4$\\ \hline
$4.0E-3$  & $1.0024E-1 \pm 5.4E-4$ & $5.898E-2\pm 5.6E-4$\\ \hline
$5.0E-3$  & $1.4327E-1 \pm 7.3E-4$ & $7.428E-2\pm 4.4E-4$ \\ \hline
$6.0E-3$  & $1.954E-1 \pm 1.6E-3$ & --\\ \hline
$7.0E-3$  & $2.560E-1 \pm 2.4E-3$ & --\\ \hline
$8.0E-3$  & $3.272E-1 \pm 1.7E-3$ & --\\ \hline
$9.0E-3$  & $4.092E-1 \pm 3.5E-3$ & --\\ \hline
$1.0E-2$  & $5.014E-1 \pm 4.1E-3$ & $1.527E-1\pm 1.1E-3$\\ \hline
\end{tabular}
\caption{DMC energies.}
\label{table:4}
\end{table}

\end{document}